\documentclass[a4paper,fleqn,usenatbib]{mnras}

\usepackage{newtxtext,newtxmath}
\usepackage[T1]{fontenc}
\usepackage{ae,aecompl}

\usepackage{graphicx}	% Including figure files
\usepackage{amsmath}	% Advanced maths commands
\usepackage{amssymb}	% Extra maths symbols

\usepackage[usenames, dvipsnames]{color} 

\newcommand\pbar{\;|\;}
\newcommand\drv{\; \textrm{d}}

\usepackage{bm} %Doesn't work with unicode-math, use symbfit instead

\usepackage{calligra}
\DeclareMathAlphabet{\mathcalligra}{T1}{calligra}{m}{n}

\newcommand\vct[1]{\underline{#1}}

\newcommand\mtr[1]{\bm{#1}}%Doesn't work with unicode-math, use symbfit instead
\usepackage{etoolbox}
\makeatletter
\patchcmd\@combinedblfloats{\box\@outputbox}{\unvbox\@outputbox}{}{\errmessage{\noexpand patch failed}}
\makeatother

\title[GMMs for Blended Photo-$z$s]{Gaussian Mixture Models for Blended Photometric Redshifts}

\author[D. M. Jones \& A. F. Heavens]{
	Daniel M. Jones,$^{1}$\thanks{E-mail: d.jones15@imperial.ac.uk}
	Alan F. Heavens,$^{1}$
	\\
	$^{1}$Astrophysics Group \& Imperial Centre for Inference and Cosmology, Imperial College London, London SW7 2A\\
}

\date{Accepted XXX. Received YYY; in original form ZZZ}

\pubyear{2019}

\begin{document}
\label{firstpage}
\pagerange{\pageref{firstpage}--\pageref{lastpage}}
\maketitle

\begin{abstract}
Future cosmological galaxy surveys such as the Large Synoptic Survey Telescope (LSST) will photometrically observe very large numbers of galaxies. Without spectroscopy, the redshifts required for the analysis of these data will need to be inferred using photometric redshift techniques that are scalable to large sample sizes. The high number density of sources will also mean that around half are blended. We present a Bayesian photometric redshift method for blended sources that uses Gaussian mixture models to learn the joint flux-redshift distribution from a set of unblended training galaxies, and Bayesian model comparison to infer the number of galaxies comprising a blended source. The use of Gaussian mixture models renders both of these applications computationally efficient and therefore suitable for upcoming galaxy surveys.
\end{abstract}


%%%%%%%%%%%%%%%%%%%%%%%%%%%%%%%%%%%%%%%%%%%%%%%%%%

%%%%%%%%%%%%%%%%% BODY OF PAPER %%%%%%%%%%%%%%%%%%

\section{Introduction}

Photometric galaxy surveys such as the  VISTA Kilo-degree Infrared Galaxy (VIKING) survey \citep{vikingData} and the Dark Energy Survey (DES) \citep{desData} have become important probes within current observational cosmology. These surveys use photometric observations of large samples of galaxies to probe the distribution of matter in the large scale structure of the Universe. This distribution is sensitive to several phenomena of interest to cosmology such as dark energy \citep[e.g.,][]{gsDarkEnergy}, the rate of expansion described by the Hubble constant \citep[e.g.,][]{desHubble}, models beyond the standard flat $\Lambda$CDM model \citep[e.g.,][]{extendCosmo} and the sum of the neutrino masses \citep[e.g.,][]{neutrinoMass}. 

Making inferences about these phenomena requires the distribution of redshifts of galaxies in the sample. Spectroscopic observations that reach a sufficient signal-to-noise provide a way to obtain very precise redshifts. However, the size and depth of these galaxy samples render spectroscopy prohibitively time-consuming. As a result, photometric redshifts are a vital part of the analysis of cosmological galaxy surveys. 

%Galaxies are also typically separated into one of several tomographic bins~\citep[e.g.,][]{tomographyHu, tomographyConstraints, tomographyKids} to improve cosmological parameter constraints by probing how the large scale structure evolves with redshift, though a full 3D analysis is also possible \citep[e.g.,][]{3dWeakLens, 3dShearFirst, 3dShearCFHT}. 

Photometric redshift methods can broadly be characterised into two types; template-based and machine learning methods. Template-based methods parametrise the relation between flux and redshift through a set of spectral templates. Galaxy fluxes are forward modelled by redshifting these spectra and integrating over the survey filters, allowing the redshift to be inferred through standard maximum likelihood \citep[e.g.,][]{hyperz, lePhare} or Bayesian techniques \citep[e.g.,][]{bpz}. These template-based methods are easily interpretable, and Bayesian inference allows rigorous statistical uncertainties to be propagated through probability density functions (PDFs). However, their accuracy is dependent on the applicability of the template sets, which are often small\footnote{The photometric redshift software BPZ~\citep{bpz} is packaged with a set of 8 templates by default.}, to the galaxy sample of interest.

Machine learning methods, on the other hand, learn the relation between flux and redshift from a training set of galaxies with known redshifts. This relation is represented by a flexible model such as random forests~\citep[e.g., ][]{randomForest, tpz}, boosted decision trees~\citep[e.g.,][]{ArborZ}, neural networks~\citep[e.g., ][]{annz, annz2}, support vector machines~\citep[e.g.,][]{svm} and Gaussian processes~\citep[e.g.,][]{firstGPPhotoz, newGPz}. Machine learning approaches can also be extended to include extra input features such as morphology~\citep[e.g.,][]{morphoz} or to use entire images as input~\citep[e.g.,][]{imagePhotoz}, rather than reducing this information to a vector of fluxes. 

The data-driven approach of machine learning methods avoids the potential pitfalls of small template sets, but instead relies on the training set being representative. If this is the case, the accuracy of these methods can be greater than that of than template-based methods~\citep{photozAccuracy} In practice however, training sets are often shallower than the photometric sample, reducing the accuracy of machine learning methods~\citep{mlRedshiftShallow}.

The unrepresentativeness of training sets could be a problem for future surveys such as the Large Synoptic Survey Telescope \citep[LSST;][]{lsstSummary} because their photometry will reach depths beyond which spectroscopy can be reasonably performed. However, by increasing the number density of galaxies on the sky, these very deep observations enable cosmological constraints with higher precision than current galaxy surveys.

Another major challenge presented by the depth of future surveys like LSST is blending \citep{lsstDensity}, the chance overlapping of galaxies along the line of sight. As a result, the intrinsic fluxes of blended galaxies cannot be observed directly, only their noisy combination. The focus of this paper is to address the challenge of inferring redshifts from this blended photometry in a way that can scale to the large datasets of future surveys.

One approach to counter the problem of blending is to separate blended sources into distinct images of each constituent galaxy, known as deblending. Deblending methods that rely solely on the morphological information contained within a single band \citep[e.g.,][]{sdssDeblend} will separate blended galaxies with large angular separations more easily than those that are more closely aligned. \cite{lsstBlendReport} found that for a survey like LSST where $44-55\%$ of sources are blended, these methods would misidentify $15-20\%$ of all sources as unblended. As a result, more recent deblending methods \citep[e.g.,][]{muscadet, scarlet} also utilise colour information by using galaxy images in several bands.

When splitting the analysis into separate images in this way, it is important to take care with how uncertainties from the deblending process are propagated. While the total flux of a source may be well constrained by observations, the separate flux of each galaxy is not, as it is not observed independently of other galaxies it is blended with. As a result, the errors on the fluxes of each galaxy will be correlated. Ideally, this correlation should be propagated to later analyses, though these uncertainties can be difficult to estimate and propagate for these deblending methods~\citep[see, e.g.,][]{scarlet}.

An alternative to deblending is to infer quantities of interest, such as photometric redshifts, from blended data directly. This joint approach automatically accounts for correlations between each galaxy in a blended source and correctly propagates these uncertainties to the final results. This is the approach taken in \cite{blendz}, which generalises Bayesian template-based photometric redshift methods to the case of blended observations. 

This paper takes the same joint-inference approach, but uses a Gaussian mixture model to learn the flux-redshift relation from a training set of galaxies with known redshifts. We then use this model as a prior to derive the posteriors and marginal-likelihoods for sources consisting of one or two galaxies. Since these can be computed analytically, this is significantly less computationally demanding than the nested sampling-based method described in \cite{blendz}, an important property for use in future galaxy surveys.

{Gaussian mixture models have previously been used for obtaining photometric redshifts of quasars~\cite{xdQSOz}. The method presented in this paper is an extension of this approach to deriving posterior distributions and Bayesian evidences for the redshifts of blended sources. This also builds on our previous work~\citep{blendz} by being completely data-driven, learning the mapping between flux and redshift from a training set, rather than imposing it \textit{a priori} through a set of templates. The significant computational advantages afforded by this approach now allows blended photometric redshifts to be applied to very large future datasets.}

Throughout this paper, we use the term \textit{constituent} to describe the individual galaxies comprising a blended source. Following convention, we refer to each multivariate Gaussian distribution in the mixture model as a \textit{component}. We denote scalars using an italic font $x$, vectors using an underlined italic font $\vct{x}$ and matrices using a bold italic font $\mtr{x}$. We summarise our notation in Table~\ref{tab:notation}.

%\begin{table*} %Changed from 2 column table to 1
\begin{table}
	\caption{A summary of the notation used throughout this paper.}
	\centering
	\label{tab:notation}
	%\begin{tabular}{p{0.15\textwidth}p{0.75\textwidth}}
	\begin{tabular}{p{0.2\columnwidth}p{0.7\columnwidth}}
		%\begin{tabular}{ll}
		\hline
		Symbol & Description \\
		\hline
		%%%%%%%%%%%%%%%%%%%%%%%%%%%%%%%%%%%%%%
		$N$	&  Number of constituent galaxies in a source \\
		%%%%%%%%%%%%%%%%%%%%%%%%%%%%%%%%%%%%%%
		$z_n$	&  Model redshift of constituent galaxy  $n$ \\
		%%%%%%%%%%%%%%%%%%%%%%%%%%%%%%%%%%%%%%
		$\vct{F}_n$	&  Model flux vector of constituent galaxy  $n$ \\
		%%%%%%%%%%%%%%%%%%%%%%%%%%%%%%%%%%%%%%
		$\hat{\vct{F}}$	&  Vector of observed fluxes \\
		%%%%%%%%%%%%%%%%%%%%%%%%%%%%%%%%%%%%%%
		$\mtr{\Sigma}^{\hat F}$	&  Covariance matrix of observed fluxes \\
		%%%%%%%%%%%%%%%%%%%%%%%%%%%%%%%%%%%%%%
		$M$	&  Number of components in the mixture model \\
		%%%%%%%%%%%%%%%%%%%%%%%%%%%%%%%%%%%%%%
		$w^k$ & Weight of mixture component $k$ \\
		%%%%%%%%%%%%%%%%%%%%%%%%%%%%%%%%%%%%%%
		$\vct{\mu}^k$ & Mean vector of mixture component $k$ \\
		%%%%%%%%%%%%%%%%%%%%%%%%%%%%%%%%%%%%%%
		$\mtr{\Sigma}^k$ & Covariance matrix of mixture component $k$ \\
		%%%%%%%%%%%%%%%%%%%%%%%%%%%%%%%%%%%%%%		
		$\mathcal{E}^1$ & Evidence for single-constituent model \\
		%%%%%%%%%%%%%%%%%%%%%%%%%%%%%%%%%%%%%%		
		$\mathcal{E}^2$ & Evidence for two-constituent model \\
		%%%%%%%%%%%%%%%%%%%%%%%%%%%%%%%%%%%%%%		
		$\mathcal{N} (\vct{x} \pbar \vct{\mu}, \mtr{\Sigma})$ & Multivariate Gaussian PDF with mean vector $\vct{\mu}$ and covariance matrix $\mtr{\Sigma}$\\		
		%%%%%%%%%%%%%%%%%%%%%%%%%%%%%%%%%%%%%%		
		$\tilde{\mathcal{N}}(\vct{x} | \vct{\eta}, \mtr{\Lambda})$ & Multivariate Gaussian PDF in natural parametrisation with parameters	$\mtr{\Lambda} \equiv \mtr{\Sigma}^{-1}$ and $\vct{\eta} \equiv \mtr{\Sigma}^{-1} \vct{\mu}$\\	
		%%%%%%%%%%%%%%%%%%%%%%%%%%%%%%%%%%%%%%	
		
		\hline
	\end{tabular}
\end{table}
%\end{table*}

In section~\ref{sec:blendz}, we briefly describe the results of \cite{blendz} and introduce our blended photo-z formalism. In section~\ref{sec:gmm}, we introduce our formalism for blended photometric redshifts with Gaussian mixture models. {We use this to derive expressions for the posteriors and evidences in section~\ref{sec:gmm-deriving}.} We present results of tests of our method on simulated data in section~\ref{sec:sim-results}. Finally, in section~\ref{sec:gama-results}, we present these tests on real blended data from the Galaxy And Mass Assembly (GAMA) survey~\citep{gamaData}.

\section{Blended photometric redshifts} \label{sec:blendz}

A Bayesian method for inferring the photometric redshifts of blended sources was introduced in \cite{blendz}.
This is a template-based method, generalising the commonly used Bayesian Photo-z (BPZ) method of \cite{bpz}. A summary of the main result, the joint posterior distribution of the redshift and magnitude of each galaxy within a blended source, is given below.

For a given template $t$ at redshift $z$, the model flux $T_{t, b}(z)$ in each band $b$ can be calculated by integrating the redshifted galaxy spectrum over the respective filter response. These fluxes are then scaled such that the flux in an arbitrarily chosen reference band $b_0$ is equal to $10^{-0.4 m_0}$, where the magnitude $m_0$ is a parameter to be inferred in addition to the redshift. The predicted flux for a blended source is then given as a linear combination of these galaxy fluxes, i.e.,
\begin{equation}
\label{eqn:blend-flux-model}
F^{(N)}_{ \{t\}, b } \big(\{z\}, \{m_0\} \big) =  \sum_{\alpha=1}^{N} \frac{10^{-0.4 m_{0, \alpha}}}{ T_{t_\alpha, b_0} \big(z_\alpha \big)} T_{t_\alpha, b} \big(z_\alpha \big) \,,
\end{equation}
where $z_\alpha$, $m_{0, \alpha}$ and $t_\alpha$ are the redshift, reference-band magnitude and template for constituent $\alpha$ respectively, and $N$ is the number of constituents in the source. The desired posterior can then be found by marginalising over the template for each galaxy and applying Bayes rule to give
\begin{equation}
\label{eqn:blendz-posterior}
\begin{aligned}
& P\Big(\{z\}, \{m_0\}  \,\Big|\, \hat{\vct{F}}, N\Big) 
\propto
\\ & \indent 
\sum_{i=1}^{T^{N}} 
P \Big(\hat{\vct{F}}, \,\Big|\, \{z\}, \{t\}_i, \{m_0\} , N \Big) 
P \Big(\{z\}, \{t\}_i, \{m_0\}  \,\Big|\, N \Big) \,. 
\end{aligned}
\end{equation}
where $\hat{\vct{F}}$ is the vector of observed fluxes and $T$ is the number of templates in the template set. The predicted flux in equation~\ref{eqn:blend-flux-model} is defined for a particular choice of template for each galaxy within the blended source. The template marginalisation therefore runs over the $T^N$ combinations of this choice. Note that this posterior is conditioned on a particular choice of $N$, the number of galaxies within the blended source; setting this is described in section~\ref{sec:model-selection}.

The joint prior can then be developed by factorising into priors defined for each constituent. In doing this, three blending-specific complications arise. Firstly, the redshifts of each constituent are not independent since galaxies are clustered. As a result, the posterior for $N$ blended sources should include an additional term involving correlation functions up to $N$-point to account for this. 

Secondly, the effect of source selection should also be accounted for. One effect of this is that the selection criteria imposes a faint-end cut on the magnitude prior. Without this cut, its simple analytic form would be improper, rendering the model selection described in section~\ref{sec:model-selection} impossible.

Lastly, a sorting condition is required to break the exchangeability of the blended constituents. Allowing this exchangeability results in marginal redshift distributions with contributions from every constituent, i.e., they would always have multiple peaks. By enforcing an ordering, the posterior better represents the underlying physical source. \cite{blendz} found that redshifts were recovered more successfully when applying this sorting condition to constituent redshifts, though sorting the magnitudes is also sufficient to break the exchangeability.

\subsection{Model selection for identifying blends} \label{sec:model-selection}

In addition to inferring the redshift of each constituent in a blended source, the method of \cite{blendz} can also identify whether a source is blended. Since the posterior defined in equation~\ref{eqn:blendz-posterior} is conditioned on the number of constituents $N$, we can consider this choice to be the model and use Bayesian model comparison techniques to infer the number of constituents within the source. 

To compare two models with a source of $n$ and $m$ constituents, we write the relative probability and apply Bayes rule to give
\begin{equation}
\begin{aligned}
\label{eqn:model-select}
\mathcal{P}_{n,m} &= \frac{P \Big( N = n\,\Big|\,  \hat{\vct{F}}, \hat F_0  \Big)}{P \Big( N = m\,\Big|\,  \hat{\vct{F}}, \hat F_0  \Big)}  
= 
\frac{P \Big(N = n  \Big)}{P \Big( N = m \Big)}
\frac{P \Big( \hat{\vct{F}}, \hat F_0 \,\Big|\, N = n  \Big)}{P \Big(  \hat{\vct{F}}, \hat F_0 \,\Big|\,  N = m\Big)}
 \,.
 \end{aligned}
\end{equation}
The first term is the ratio of model priors, allowing the \textit{a priori} probability of a source being blended to be set. This value could be informed by the expected number of blended sources given the survey depth, or could leverage additional independent information such as whether a source is located within a cluster or the field. Throughout, we assume this ratio is unity, so that a source is equally likely to be blended as not. This assumption is trivial to modify, however.

The second term in equation~\ref{eqn:model-select} is a ratio of marginal likelihoods known as the Bayes factor. The calculation of these marginal likelihoods, also known as evidences and labelled $\mathcal{E}$, involves an integral over the full support of the prior, i.e., 
\begin{equation}
\label{eqn:evidence-integral}
\mathcal{E} \equiv P \Big(\vct{d} \,\Big|\, \mathcal{M} \Big) = \int P \Big(\vct{d} \,\Big|\,\{\theta\}, \mathcal{M} \Big) P \Big(\{\theta\} \,\Big|\, \mathcal{M} \Big) \textrm{d} \{\theta\} \,,
\end{equation}
where $\vct{d}$ is the data vector, $\mathcal{M}$ is the model and the integral is over the set of model parameters $\{\theta\}$. 

This integral is often difficult to evaluate, particularly if the dimensionality of the parameter space is large. While the prior volume may be large, the likelihood can peak sharply. Nevertheless, the comparatively low-density tails of the posterior can contain significant volume and can therefore not be ignored. Numerically evaluating an integral with non-negligible contributions at both of these scales is computationally challenging.

%One method to evaluate the evidence is nested sampling~\citep{nestSamp}. This proceeds by repeatedly sampling within contours such that each sample is at a higher likelihood than the previous sample. Each successive contour thus encloses a smaller prior volume than the previous. By doing this, nested sampling transforms the integral of equation~\ref{eqn:evidence-integral} into one consisting of a single dimension only, the likelihood as a function of the prior volume enclosed by its contour. This integrand is also smooth and monotonically decreasing, vastly simplifying the numerical integration at the expense of complicating the sampling procedure. By weighting appropriately, these samples can also be used for parameter inference.

{In order to sample the posterior and evaluate the evidence, \cite{blendz} uses MultiNest~\citep{multinest}, an efficient implementation of the nested sampling method~\citep{nestSamp}.} 
%an efficient way to draw samples required for nested sampling by approximating the threshold likelihood contour with a series of ellipses. 
%This is the method used in \cite{blendz}. 
However, even sampling with an efficient method such as MultiNest can be computationally demanding; sampling both the two- and four-dimensional posteriors for one- and two-constituent sources respectively\footnote{Both a redshift and a magnitude is sampled for each constituent in the source.} takes approximately two minutes per source on a workstation with a $3\,\textrm{GHz}$ Intel Xeon processor. While this is viable for small samples, it is not scalable to the large samples of $\sim 10^9$ galaxies in a future survey like LSST. Instead, this paper develops a method that does not rely on these computationally demanding integrals. As a result, the one- and two-constituent inference and model selection can be done for approximately ten sources per second, a speed-up of three orders of magnitude on the workstation described above. Photometric redshift inference is also trivially parallelisable for high-performance computing environments, since each source can be considered independently. We present our method below.

\section{Gaussian mixture model photo-z} \label{sec:gmm}

Photometric redshifts inferred using machine learning methods are often very accurate when good training data is available. These methods perform regression, and use this training data to learn the mapping from fluxes to redshifts. Many machine learning algorithms are not inherently probabilistic; a particular input will map to a particular output. However, accurate uncertainties on cosmological parameters rely on propagating uncertainties from all stages of the analysis. Machine learning photometric redshift methods have therefore developed several ways to estimate these uncertainties. 

One example that accounts for errors in the observed fluxes is to apply the chain rule to successive layers of a neural network~\citep{annz}, providing the variance of the output redshift. Some machine learning methods such as a Gaussian process \citep[e.g.,][]{firstGPPhotoz}, are already explicitly probabilistic, naturally producing variance estimates alongside their prediction. Other methods can represent their uncertainties more generally by inferring PDFs as their output. This can be done by training many machine learning algorithms to each independently estimate the redshift and taking the distribution of the ensemble to be the redshift PDF \citep{annz2}. A single neural network can also accomplish this by being trained to output the parameters of a parametrised PDF rather than the redshift directly \citep{imagePhotoz}. {PDFs represent the complete probabilistic knowledge over a system under investigation, and are thus a general mechanism for quantifying and propagating uncertainties within a statistical analysis~\citep[e.g.][]{redBook}.}

In addition to enabling the rigorous propagation of uncertainties, using full photometric redshift PDFs has been shown to improve the accuracy of cosmological analyses \citep[e.g.,][]{wlWithPDF, clusterWithPDF}. PDFs also have an advantage over simply representing uncertainty with the variance in their ability to represent multimodality; that is, several distinct, well separated redshifts being plausible for a given vector of fluxes. This is a common occurrence in photometric redshifts~\citep{photozFilters}. Colour-redshift degeneracies mean that high- and low- redshift galaxies can have similar colours, often due to spectral features such as the Lyman and Balmer breaks being misidentified as one another~\citep{lymanBalmer}.

Here, we treat the training data not as variables to regress between, but instead as noisy samples from the joint redshift-flux distribution, turning the problem into one of density estimation. The joint density is the most general probabilistic description of the training data, allowing several quantities of interest to be derived. Given an observed vector of fluxes $\hat{\vct F}$, the redshift can be inferred using the conditional distribution $P(z \pbar \hat{\vct F})$ which can be derived from the joint distribution. This PDF can be multimodal, capturing the degeneracy described above. These distributions can be composed together to produce the conditional distribution of the redshifts of a blended source $P(z_1, z_2 \pbar \hat{\vct F})$ in a similar fashion. The joint distribution also permits calculation of marginal likelihoods, allowing Bayesian model selection techniques to be used to infer the number of constituents in a source. Finally, the interpretation of the joint distribution is clear, in contrast to other machine learning methods that can be  `black-boxes', requiring additional ad-hoc techniques to improve their interpretability ~\cite[e.g.,][]{blackBox1, blackBox2}

We model the joint distribution of the latent, noise-free parameters as a Gaussian mixture model (GMM), a weighted linear combination of multivariate Gaussians, i.e.,
\begin{equation}
P(z, \vct{F}) = \sum_k w^k \; \mathcal{N} (z, \vct{F} \pbar \vct{\mu}^k, \mtr{\Sigma}^k) \,.
\end{equation}
By imposing that $\sum_k w^k=1$, this density is correctly normalised, i.e., 
\begin{equation}
\label{eqn:prior-normalised}
\sum_k w^k \int \!\!\! \int \mathcal{N} (z, \vct{F} \pbar \vct{\mu}^k, \mtr{\Sigma}^k) \drv z \drv \vct{F} = \sum_k w^k  =1 \,.
\end{equation}

\begin{figure*}
	\includegraphics[width=\textwidth]{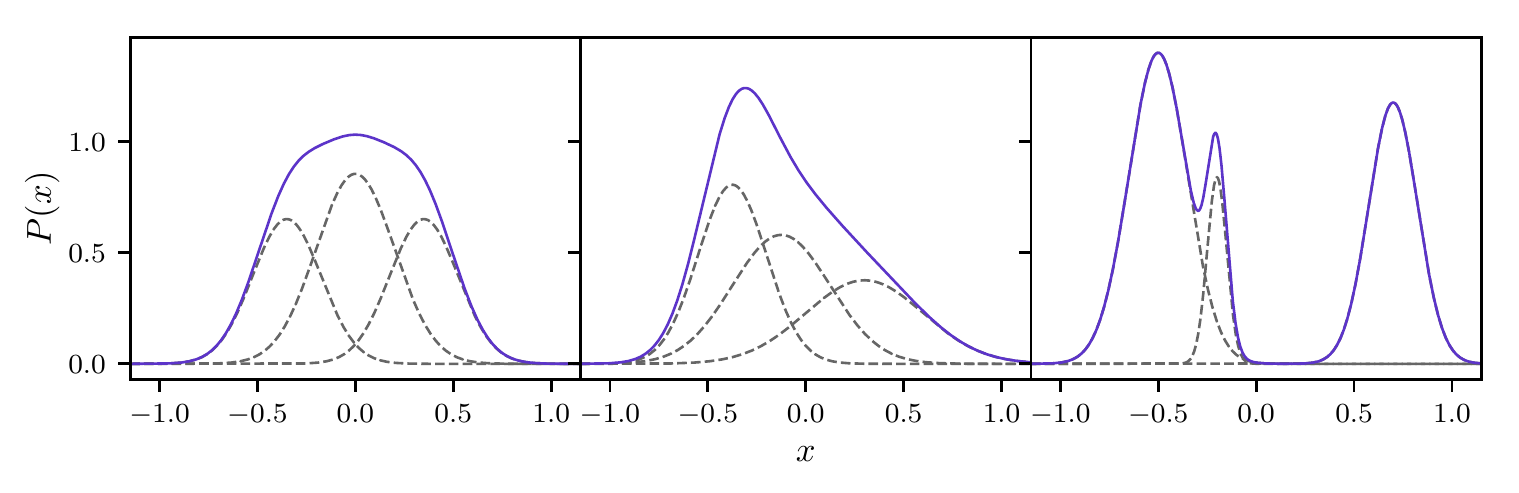}
	\caption{Plot showing a variety of PDFs that can be represented by Gaussian mixture models, given a sufficient number of components. The dashed grey curves show each weighted Gaussian component, and the solid blue curves show the mixture formed by the linear combination of these components.}
	\label{fig:toy-gmm-example}
\end{figure*}

This choice has several useful features. Firstly, GMMs are easy to train using standard, well-tested methods. This is discussed further in section~\ref{sec:gmm-train}. Secondly, inference with GMMs is computationally inexpensive as they can be efficiently sampled as detailed in section~\ref{sec:gmm-inference}. Lastly, GMMs are mathematically convenient. Both the conditional and marginal distributions of multivariate Gaussians are also Gaussians. The same is also true of both the product and convolution of several multivariate Gaussians. These properties will be used frequently throughout this paper to render many calculations analytic. Despite this, GMMs can represent a wide variety of PDFs, including those that are skewed or multimodal. This is demonstrated in Fig.~\ref{fig:toy-gmm-example}.

Using GMMs to infer photometric redshifts in this way was first done in \cite{xdQSOz}, who applied the method to obtain photometric redshifts of quasars and used model selection techniques to separate stars and quasars. The method we present in this paper differs from this in several ways. Firstly, we extend the method to the case of jointly inferring multiple redshifts directly from blended data.  

Secondly, \cite{xdQSOz} fit a series of many GMMs to the fluxes and redshifts of quasars in several magnitude bins. As a result, our model has significantly fewer parameters to fit. Nevertheless, our use of cross-validation to set the number of mixture components as described in section~\ref{sec:gmm-cross-val} provides the model sufficient flexibility to fit the flux-redshift density with the full fidelity provided by the training set.

{The binning of \cite{xdQSOz} is not possible due to the extension to blended sources. Observations in this case are of the flux of the blended source, while the magnitude bin in that model is chosen based on the magnitude of an individual galaxy. This quantity that is not observed in the blended case, and so cannot be used to choose a magnitude bin. The same is true of colours, i.e., ratios of fluxes relative to the flux in a particular reference band, which are often used in machine learning-based photometric redshift methods. Since the reference-band flux of each galaxy in a blended source is not observed, the colours for each galaxy cannot be calculated and so cannot be used to infer the redshifts.}

Finally, our derivation does not use the  convolution property of multivariate Gaussians described above, since integrals over fluxes are then implicitly evaluated from $-\infty$ to $\infty$ as multivariate Gaussians have infinite support. These integrals therefore contain contributions from non-physical negative fluxes. This is a safe approximation when considering unblended sources, since their flux is strongly constrained by observations. However, the same is not true of blended sources, where the individual flux of each constituent is not observed. Instead, we evaluate these results using an efficient Monte Carlo integration method.  We therefore treat unblended sources in the same way for consistency.

All fluxes throughout are renormalised for numerical stability. This is done by dividing each flux by the standard deviation in the training set, e.g., for band $b$,
\begin{equation}
F_b \rightarrow \frac{F_b}{\sigma(\{ \hat{\vct F_b} \}_{\textrm{tr}})}  \,.
\end{equation}
{Normalising the data in this way is a common preprocessing step in machine learning methods. Without this renormalisation, the observed fluxes are small enough that the EM fitting procedure is dominated by numerical errors as the covariance matrices of the components become poorly conditioned.}
The corresponding change in the covariance matrix of each data point is given by
\begin{equation}
{\Sigma}_{ij} \rightarrow \frac{{\Sigma}_{ij}}{\sigma(\{ \hat{\vct F_i} \}_{\textrm{tr}}) \sigma(\{ \hat{\vct F_j} \}_{\textrm{tr}}) } \,.
\end{equation}

{We also note that magnitudes are commonly used for this purpose in machine learning-based photometric redshift methods, since the logarithmic transformation of the flux also effectively normalises them. However, an advantage of the GMM method presented here is that expressions for posteriors and evidences can be calculated analytically. This relies on the model for the flux of the blended sources being a linear combination of the fluxes of the individual constituents, since this leaves the likelihood of the sum a Gaussian. This would no longer be the case when using magnitudes, as the model for the magnitude of a blended source would be a non-linear function of the individual constituent magnitudes.}

\subsection{Training Gaussian mixture models} \label{sec:gmm-train}

Our prior density $P(z, \vct{F})$ is defined in terms of the true, latent parameters. Therefore, this density must be fitted with a method that incorporates both the noisy data and the covariance. {To do this, we use} extreme deconvolution \citep{extremeDeconv}, an extension of the expectation-maximisation (EM) algorithm~\citep{expectMax} commonly used the find the maximum-likelihood parameters of GMMs. {This is the same fitting method as the quasar photometric redshift method of \cite{xdQSOz}.} 

Extreme deconvolution generalises the EM algorithm to the case where the data is subject to normally-distributed errors.
The EM algorithm is a general method for fitting models with some form of hidden data in addition to the observed data. Given an initial guess at the parameters, the algorithm iteratively modifies these parameters to increase the likelihood, converging to a local maximum. 

For a single multivariate Gaussian, the maximum-likelihood parameters can be found exactly through the derivative of the likelihood. However, the same is not true of mixtures of Gaussians, as these parameters are not available in closed form. The hidden information that would make this tractable is the identity of the component from which each sample was drawn. If this were known, fitting the GMM would reduce to the previous analytic case. Though this information is hidden, this points to an iterative solution; first, the parameter guess can be used to update the hidden information, then this information can be used to update the parameters.

In essence, expectation-maximisation is a probabilistic version of this procedure that takes into account the uncertainty in the hidden information. By averaging the likelihood over the probability of each sample being drawn from each component, the maximum likelihood parameters can be found in closed form. Since the component probability depends on the parameters being fitted, this process is iterative.

The extreme deconvolution method of \cite{extremeDeconv} extends the EM algorithm to fit data with Gaussian errors. This is done by replacing the likelihood with a marginalised version given by
\begin{equation}
\begin{aligned}
P(\hat{\vct x} \pbar \{\theta\}) &=
\int P(\hat{\vct x}, \vct{x} \pbar \{\theta\}) \drv \vct{x}
=
\int 
P(\hat{\vct x} \pbar \vct{x}) 
P(\vct{x} \pbar \{\theta\})
\drv \vct{x} \,,
\end{aligned}
\end{equation}
where $\hat{\vct x}$ is the vector of observed values, $\vct{x}$ is the latent vector of true values and $\{\theta\}$ are the mixture parameters being fitted, i.e., weights, means and covariances. The data likelihood $P(\hat{\vct x} \pbar \vct{x})$ is assumed to be a multivariate Gaussian, and $P(\vct{x} \pbar \{\theta\})$ is the GMM. Due to the convolution property of multivariate Gaussians, this marginalised likelihood is also a Gaussian mixture, and thus amenable to being fitted using an expectation-maximisation approach. {Using this extreme deconvolution method, we fit the joint flux-redshift distribution $P(z, \vct{F})$ while accounting for uncertainties in the training set.}

This fitting procedure assumes that the number of mixture components is fixed. The method we use to decide on this number is discussed in section~\ref{sec:gmm-cross-val}.

As discussed above, multivariate Gaussians have infinite support, and so non-physical negative fluxes and negative redshifts are \textit{a priori} allowed. No non-physical fluxes will be present in the training set, and negative redshifts, while not non-physical, are sufficiently rare that they can be presumed to not be present either. As a result, there is no incentive for the training algorithm to assign significant prior volume here. However, without an additional prior on the mixture parameters, prior volume in negative regions is not penalised either. 

It is possible to generalise {EM-based methods such as extreme deconvolution} to maximise the posterior rather than the likelihood by adding a log-prior. However, while this will ameliorate the problem of negative values, it cannot eliminate it completely; the GMM having infinite support means that every point in parameter space will always have non-zero density.

An alternative approach is to impose an additional prior that is zero is any negative regions of parameter space, i.e.,
\begin{equation}
P(z, \vct{F}) = \Theta(z, \vct{F}) \sum_k w^k  \, \mathcal{N} (z, \vct{F} \pbar \vct{\mu}^k, \mtr{\Sigma}^k) \,\
\end{equation}
where 
\begin{equation}
\label{eqn:phys-prior}
\Theta(z, \vct{F}) = 
\begin{cases}
0 \quad \textrm{for } z, \vct{F} < 0
\\
1 \quad \textrm{otherwise.}
\end{cases}
\end{equation}

This will exactly fix the problem of negative values. However, it will also force otherwise analytic integrations to have to be done numerically. These cases are discussed in the relevant sections below.

Imposing this boundary prior will also change the normalisation of the prior from unity, i.e., 
\begin{equation}
\int \!\!\! \int \Theta(z, \vct{F}) \sum_k w^k  \, \mathcal{N} (z, \vct{F} \pbar \vct{\mu}^k, \mtr{\Sigma}^k) \drv z \drv \vct{F} \neq 1
\,.
\end{equation}
The model selection described below requires that the prior be normalised. This normalisation differs between the single- and two-constituent cases, with the latter also being affected by the sorting condition. These normalisations are therefore discussed in their respective sections below.

It should be noted that, since this is an empirical method that does not rely on any underlying physical model in the way that a template-based method does, the redshift can be transformed almost arbitrarily. The only restrictions in this transformation are that it is both invertible and well-defined for all positive real numbers. The only modifications to the method required to accommodate this are to the limits of redshift integrals. For a transformation $\mathcal{T}(z)$, the lower and upper limits should be replaced with $\mathcal{T}(0)$ and $\mathcal{T}(\infty)$ respectively. 

The transformation $\mathcal{T}(z) = \log(z)$ would seem to be a sensible choice, as the lower and upper integration limits would become $-\infty$ and $\infty$ respectively, rendering all the redshift integrations throughout analytic. This is the approach taken by \cite{xdQSOz}. However, in our tests, we found that this transformation reduces the accuracy of the blended redshift inference. The difference in accuracy of the single redshift inference was negligible. As a result, we do not transform redshifts throughout this paper.

\begin{figure*}
 	\includegraphics[width=\textwidth]{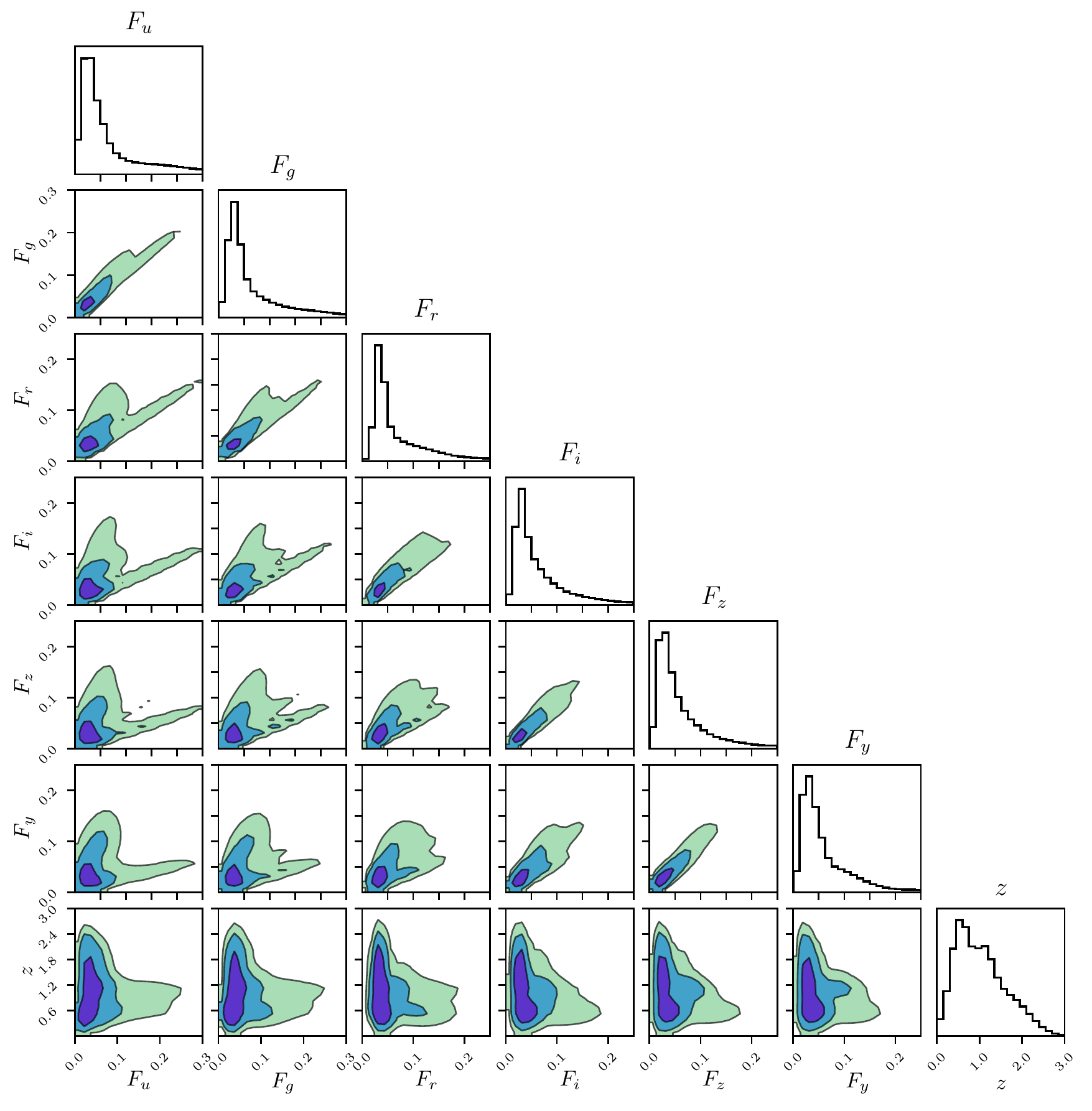}
 	\caption{Corner plot of an example flux-redshift distribution fitted by our model. This density shown here is visualised using $10^6$ samples drawn from a model that was fitted to the LSST-like simulations presented in section~\ref{sec:sim-results}.}
	\label{fig:gmm-prior-corner}
 \end{figure*}

 A plot of this prior distribution, fitted to the simulated LSST-like training data described in section~\ref{sec:sim-results} and plotted using \texttt{corner.py}~\citep{corner}, is shown in Fig.~\ref{fig:gmm-prior-corner}. {The ability to plot this distribution is an advantage to this GMM method. As described above, machine learning methods can act as black boxes, where what has been learned is a complicated function approximator that can be difficult to interpret. In contrast, the central object being learned here is the joint flux-redshift distribution, a meaningful statistical object that can be plotted, sampled from and manipulated mathematically.}
 
\subsection{Utilising blended training data}

The derivations detailed in sections~\ref{sec:gmm-single-post} and \ref{sec:gmm-single-evd} are presented for a scalar redshift $z$. However, it should be noted that these single-constituent results also hold for a vector $\vct{z}$. As a result, this method can be generalised so that the model is fitted to blended training data, i.e., a vector of blended fluxes with the associated vector of redshifts for each constituent.

Utilising blended training data would allow the method to infer both redshifts and the number of constituents accurately in cases where the blended constituents were systematically different from non-blended constituents. The cost of this, however, is an increase in the required size of the training set. Machine learning-based methods require a training set that is representative of the test set in order to be accurate. A blended training set would therefore have to contain sufficient examples of all possible pairs of constituents, rather than the constituents alone as required for the results in sections~\ref{sec:gmm-blend-post} and \ref{sec:gmm-blend-evd}.

\subsection{Cross-validating the number of mixture components} \label{sec:gmm-cross-val}

The procedure described in section~\ref{sec:gmm-train} will fit the weights, means and covariances of the GMM for a fixed number of components. However, it is difficult \textit{a priori} to choose this number; including more components within the mixture allows it more flexibility, but too many will cause the model to overfit. Given enough mixture components, the variance of each component will approach zero, with each  being responsible for only a single sample. While this will significantly increase the likelihood of the training set, it will also cause the model to generalise extremely poorly. 
%Side note - this is a general thing in ML, local maxima are fine because the global maximum is massively overfit!

Overfitting is a general concern when fitting machine learning models. As a result, various techniques for preventing overfitting have been suggested. These include restricting the dimensionality of the parameter space as we do here by fixing the number of components, disfavouring overfitted parameters through regularisation \cite[e.g.,][]{ridge} or Bayesian priors \cite[e.g.,][]{bnn}, and stopping training before overfitting occurs \citep[e.g.,][]{earlyStop}. 

The ability for a machine learning method to generalise and whether it has been overfitted can be tested by using a a validation set, an additional set of data where the input and output are known but is not used during the training. By measuring the difference between the prediction and the known ground truth, the model can be evaluated.

It is useful to point out that a corollary to the notion of overfitting is that the fitting procedure need not converge to a global maximum, as that set of parameters will overfit the data. Instead, local maxima can be nearly as accurate on the test set, while generalising much better \citep{lossSurface}. Therefore, it is reasonable to use parameters corresponding to local maxima that are found to perform well during validation. This can avoid expending significant optimisation effort attempting to fit the global maximum.

To choose the number of components, we use k-fold cross validation, a method that repeatedly splits the data into training and validation sets. The training set is first split into $k$ subsets. The model is then trained on $k-1$ subsets of this data, assuming a fixed number of mixture components $M$. The remaining subset is then used for validation. By evaluating the model using the fluxes of this subset, the redshift predictions can be compared to the known truth and scored based on their accuracy. This training and validation is repeated $k$ times for each number of components considered so that each subset is used for evaluation once. The average score can then be used to evaluate each number of components. 

To evaluate the accuracy of the redshift predictions, we use the RMS scatter. Given a predicted redshift $z_{\textrm{p}, g}$ and a  spectroscopic redshift $\hat z_{\textrm{s}, g}$ for galaxy $g$, the normalised error is defined as
\begin{equation}
\label{eqn:norm-error}
\tilde{\delta}_g = \frac{\hat z_{\textrm{s}, g} - z_{\textrm{p}, g}}{1 + \hat z_{\textrm{s}, g}} \,.
\end{equation}
After calculating this error for $n_\textrm{g}$ galaxies, the RMS scatter for the sample is then given by
\begin{equation}
\label{eqn:sig-rms}
\sigma_{\textrm{RMS}} =  \sqrt{\frac{1}{n_\textrm{g}} \sum_g \tilde{\delta}_g^2  }\,.
\end{equation}
This metric is evaluated using $k$-fold validation for each number of mixture components $M$ being considered. We then choose $M$ to be the number of components that minimises the RMS scatter averaged over each of the $k$ folds.

\subsection{Sampling from Gaussian mixture models}\label{sec:gmm-inference}

One of the significant advantages of using GMMs is that they can be efficiently sampled from without using methods such as MCMC. Since they are simply linear combinations of component distributions, a simple sampling scheme is to randomly select one of the components with a probability given by the weights, and then to draw a sample from the respective multivariate Gaussian. 
%This approach can be generalised to $N$ samples by sampling a vector of counts; that is, the number of times each component would be chosen to be sampled from, given the weights. These counts are distributed with a multinomial distribution and thus can be efficiently sampled. Firstly, the weights are used to define bins on the range $[0, 1]$ so that bin $k$ is of size $w_k$, i.e., the lower and upper boundaries of bin $k$ are given by $\sum_{i=1}^{k-1} w_i$ and $\sum_{i=1}^{k} w_i$ respectively. $N$ uniform random numbers $U \in [0, 1]$ are then drawn from a pseudo-random number generator \citep[e.g.,][]{mersenneTwist}. Finally, the count of component $k$ is incremented by one for each sample from $U$ within bin $k$.

%Once each sample has been assigned a multivariate Gaussian component, this distribution can then be sampled as follows. Given the Cholesky decomposition of the covariance matrix $\mtr{\Sigma} = \mtr{L}\mtr{L}^T$ where $\mtr{L}$ is a lower-triangular matrix, random samples $\vct{x}$ drawn from a standard multivariate Gaussian $\mathcal{N}(\vct{x} \pbar \vct{0}, \mtr{I})$ can be transformed to ones drawn from any general multivariate Gaussian $\mathcal{N}(\vct{y} \pbar \vct{\mu}, \mtr{\Sigma})$ by $\vct{y} = \mtr{L} \vct{x} + \vct{\mu}$. Samples from a standard multivariate Gaussian can be obtained from uniform random samples using a Box-Muller transform~\citep{boxMuller}.

This sampling scheme allows GMMs to be sampled efficiently and without rejection. However, the addition of the boundary prior described in section~\ref{sec:gmm-single-post} means that samples with negative fluxes and redshifts are rejected during inference. Nevertheless, the efficiency of this sampling scheme means that this does not pose a problem, since many samples can still be drawn from the relevant posterior with little computational effort.

\subsection{Compressed storage of PDFs} \label{sec:gmm-storage}

As described above, it is important that the results of photometric redshifts are represented as a PDF. However, given the large sample sizes of future galaxy surveys like LSST, storing these PDFs can present a problem. While a point estimate of the redshift and an associated error can be stored simply as two real numbers, a PDF will generally require many more. A naive representation of this distribution is a histogram where the redshift bins are fixed for all sources. While this is simple, it is not space efficient.

This problem was first investigated by \cite{compress1}, which proposed a sparse basis representation using Gaussian and Voigt distributions. Using this method, the PDF can be stored in a single signed integer per basis function, with $\mathcal{O}(10)$ basis functions required to accurately reconstruct the original PDFs. \cite{compress2} test PDF compression methods by measuring the Kullback-Leibler divergence between the original and compressed PDFs. They suggest storing the redshifts corresponding to equally-spaced quantiles as an alternative to histograms.

The posteriors presented here are GMMs, potentially multiplied by an additional physical constraint. This representation permits a simple compression technique of discarding low-weight components. By construction, the number of components in the mixture describing the prior is the same as the mixture describing the redshift posterior. However, the latter is generally significantly more compact, describing the density over the parameter space for a single source only, rather than the entire population. It is therefore reasonable to expect that this posterior distribution could be represented by fewer components than the prior. 

If additional computation can be afforded for a further reduction in storage space, mixture components can also be merged into a smaller number of approximating components. This procedure is known as mixture reduction \cite[see, e.g.,][]{reduction1, reduction2, reduction3, reduction4}.

\section{Deriving posteriors and evidences} \label{sec:gmm-deriving}

\subsection{Single-constituent posterior} \label{sec:gmm-single-post}

We now derive the posterior distribution assuming that the source consists of a single, unblended constituent galaxy. The redshift under this model can then be inferred by sampling from this posterior, as described in section~\ref{sec:gmm-inference}. We start by marginalising over the true, latent flux vector $\vct F$, giving
\begin{equation}
\begin{aligned}
P(z \pbar \hat{\vct F}) 
&= \int P(z, \vct{F} \pbar \hat{\vct F}) \drv \vct{F} \,.
\end{aligned}
\end{equation}
Applying Bayes rule, this becomes
\begin{equation}
\label{eqn:post1_bayes}
\begin{aligned}
P(z \pbar \hat{\vct F}) 
&\propto \int P(\hat{\vct F} \pbar \vct{F}) \; P(z, \vct{F}) \drv \vct{F} \,,
\end{aligned}
\end{equation}
where the unnecessary redshift conditioning has been dropped from the likelihood. We assume the likelihood to be a multivariate Gaussian centred on the observed fluxes, i.e., 
\begin{equation}
P(\hat{\vct F} \pbar \vct{F}) = \mathcal{N}(\vct{F} \pbar \hat{\vct F} , \mtr{\Sigma}^{\hat F} )
\,,
\end{equation}
where $\mtr{\Sigma}^{\hat F}$ is the covariance matrix of the observation. Galaxy surveys typically assume the errors on observed fluxes in each band to be independent, i.e., given as a flux and an error. In this case, the covariance matrix would simply be diagonal. No assumption is made about this covariance throughout however, allowing fluxes to be correlated in general. 

The prior in equation~\ref{eqn:post1_bayes} is given by the GMM described above. This prior is the only term involving the redshift; it fully represents the relation between flux and redshift learned from the training set.

Inserting both the prior and the likelihood into equation~\ref{eqn:post1_bayes}, the posterior becomes

\begin{equation}
\label{eqn:post1_P+L}
P(z \pbar \hat{\vct F}) \propto 
\sum_k w^k \int 
\mathcal{N}(\vct{F} \pbar \hat{\vct F} , \mtr{\Sigma}^{\hat F} )
\;
\mathcal{N} (z, \vct{F} \pbar \vct{\mu}^k, \mtr{\Sigma}^k)
\drv \vct{F} \,.
\end{equation}

This posterior now contains the product of two Gaussian PDFs, albeit with different dimensionalities. We proceed by combining these two densities into a single multivariate Gaussian. This is analogous to the derivation of \cite{xdQSOz}. However, as described above, we do not make use of the convolution property of multivariate Gaussians, instead forming the product explicitly. To do this, we write our posterior in terms of a parameter vector $\vct{\theta}$ partitioned into redshift and fluxes, i.e.,
\begin{equation}
\vct{\theta} = 
\left(
\begin{matrix} 
z \\
\vct{F} \\
\end{matrix}
\right) \,.
\end{equation}
Throughout, we label the redshift and flux blocks of parameters partitioned in the same way with $z$ and $f$ respectively. 

The likelihood involves only the flux partition of the parameter vector. However, our prior has support over both redshift and flux, i.e., all of $\vct{\theta}$. The component parameters are thus partitioned in the same way so that the mean and covariance are given by
\begin{equation}
\vct{\mu}^k = 
\left(
\begin{matrix} 
\mu_\textrm{z}^k \\
\vct{ \mu}_\textrm{f}^k \\
\end{matrix}
\right)
\end{equation}
and 
\begin{equation}
\mtr{\Sigma}^k = 
\left(
\begin{matrix} 
\Sigma_{\textrm{zz}}^k &&  \vct{\Sigma}_{\textrm{zf}}^k \\%&& \\
\vct{\Sigma}_{\textrm{fz}}^k &&  \mtr{\Sigma}_{\textrm{ff}}^k \\
\end{matrix}
\right) 
\end{equation}
respectively. The product of these two densities is most easily written in terms of the natural parametrisation\footnote{This is also referred to as the canonical or information parametrisation.} of the multivariate Gaussian. This has a density given by
\begin{equation}
\tilde{\mathcal{N}}(\vct{x} | \vct{\eta}, \mtr{\Lambda}) = \exp \left[ \alpha + \vct{\eta}^T \vct{x} - \frac{1}{2} \vct{x}^T \mtr{\Lambda} \vct{x} \right]\,,
\end{equation}
where we have added a tilde to notate the alternative parametrisation. The normalisation factor is given by
\begin{equation}
\alpha = - \frac{1}{2} \left[ d \log (2 \pi) - \log |\mtr{\Lambda}| + \vct{\eta}^T \mtr{\Lambda}^{-1} \vct{\eta} \right]\,,
\end{equation}
and the covariance matrix and mean vector are replaced with the natural parameters
$
\mtr{\Lambda} \equiv \mtr{\Sigma}^{-1}
$
and
$
\vct{\eta} \equiv \mtr{\Sigma}^{-1} \vct{\mu} \,.
$
respectively. The inverse covariance matrix $\mtr{\Lambda}$ is known as the precision matrix.
The product of the two densities in equation~\ref{eqn:post1_P+L} can then be combined into a single multivariate Gaussian written in this natural parametrisation, given by
\begin{equation}
\label{eqn:gmmz:combine-gauss1}
\mathcal{N}(\vct{F} \pbar \hat{\vct F} , \mtr{\Sigma}^{\hat F} )
\;
\mathcal{N} (z, \vct{F} \pbar \vct{\mu}^k, \mtr{\Sigma}^k)
=
c \tilde{\mathcal{N}}(z, \vct{F} \pbar \vct{\eta}^{k \hat F}, \mtr{\Lambda}^{k \hat F}) \,,
\end{equation}
where the new parameters are
\begin{equation}
\mtr{\Lambda}^{k \hat F} =   (\mtr{ \Sigma}^k)^{-1} +
\left(
\begin{matrix}
0 && 0\\
0 && (\mtr{\Sigma}^{\hat F})^{-1} 
\end{matrix}
\right)
\end{equation}
and
\begin{equation}
\vct{\eta}^{k \hat F} = (\mtr{\Sigma}^k)^{-1} \vct{\mu}^k +
\left(
\begin{matrix}
0 \\
(\mtr{\Sigma}^{\hat F})^{-1} \hat{\vct F}
\end{matrix}
\right) \,.
\end{equation}
Conveniently, the constant of proportionality $c$ can also be written in terms of a multivariate Gaussian in standard parametrisation. This is given by
\begin{equation}
c_1^k = \mathcal{N} (\vct{\mu}^k_{\textrm{f}} \pbar \hat{\vct F}, \mtr{\Sigma}^k_{\textrm{ff}} + \mtr{\Sigma}^{\hat F}) \,.
\end{equation}
{These results are close to a standard property~\citep[e.g., ][]{matrixCookbook} where the product of two multivariate Gaussian densities is also a multivariate Gaussian. However, the differing dimensionalities of the two densities in equation ~\ref{eqn:gmmz:combine-gauss1} slightly alter the expressions for the new parameters.}

Inserting these results into equation~\ref{eqn:post1_P+L} and moving constant terms outside of the integral, the expression for the posterior becomes
\begin{equation}
P(z \pbar \hat{\vct F}) \propto 
\sum_k w^k 
%\mathcal{N} (\vct{\mu}^k_{f} \pbar \hat{\vct F}, \mtr{\Sigma}^k_{\textrm{ff}} + \mtr{\Sigma}^{\hat F}) 
c_1^k
\int 
\tilde{\mathcal{N}}(z, \vct{F} \pbar \vct{\eta}^{k \hat F}, \mtr{\Lambda}^{k \hat F})
\drv \vct{F} \,.
\end{equation}

In principle, this integral can be done analytically by moving back to standard parametrisation, i.e., $\mtr{\Sigma}^{k \hat F} = \left(\mtr{\Lambda}^{k \hat F}\right)^{-1}$ and $\vct{\mu}^{k \hat F} = \mtr{\Sigma}^{k \hat F} \vct{\eta}^{k \hat F}$. The marginalisation can then be done by dropping the corresponding elements from the mean vector and covariance matrix, giving
\begin{equation}
\begin{aligned}
P(z \pbar \hat{\vct F}) 
&\propto \sum_k w^k 
%\mathcal{N} (\vct{\mu}^k_{f} \pbar \hat{\vct F}, \mtr{\Sigma}^k_{\textrm{ff}} + \mtr{\Sigma}^{\hat F}) 
c_1^k
\mathcal{N}(z \pbar \vct{\mu}^{k \hat F}_{\textrm{z}}, \mtr{\Sigma}^{k \hat F}_{\textrm{zz}}) \,.
\end{aligned}
\end{equation}
Note that this is simply a one-dimensional Gaussian mixture model with a new set of weights given by 
%$w^{k \hat F} \equiv w^k \; \mathcal{N} (\vct{\mu}^k_{f} \pbar \hat{\vct F}, \mtr{\Sigma}^k_{\textrm{ff}} + \mtr{\Sigma}^{\hat F})$.
$w^{k \hat F} \equiv w^k c_1^k$.

An important caveat to this result, however, is that the limits of integration are assumed to be $(-\infty, \infty)$; that is, non-physical negative fluxes contribute to the integral. This is the same assumption as used in the derivation in \cite{xdQSOz} using the convolution property of multivariate Gaussians. 
For this non-blended photo-z, this assumption is sound since the latent fluxes are strongly constrained by the likelihood, meaning that negative fluxes will be strongly down-weighted. However, this will not be the case for the blended photo-z derived in section~\ref{sec:gmm-blend-post} where only the sum of two latent flux vectors is observed. 

An alternative approach is to add the boundary prior $\Theta(z, \vct{F})$ as described in section~\ref{sec:gmm-train}. This has two effects. Firstly, the prior with this addition must be explicitly normalised, a necessary condition for the model selection. The normalisation factor is given by an integral over the unnormalised prior, i.e.,
\begin{equation}
\label{eqn:norm1}
\mathcal{A}_1 = \int \!\!\! \int \Theta(z, \vct{F}) \sum_k w^k  \, \mathcal{N} (z, \vct{F} \pbar \vct{\mu}^k, \mtr{\Sigma}^k) \drv z \drv \vct{F} \,.
\end{equation}
This integral can be efficiently estimated using Monte Carlo integration. First, a set of redshifts and fluxes $\{z, \vct{F}\}$ is sampled from the mixture, as described in section~\ref{sec:gmm-inference}. Since the prior without the boundary prior is normalised to unity as in equation~\ref{eqn:prior-normalised}, this integral is then equal to fraction of these samples obeying the boundary prior, i.e., where $\Theta(z, \vct{F}) = 1$.

The second effect of adding the boundary prior is that marginalising over fluxes is no longer analytic.
{Inserting the boundary prior and the corresponding prior normalisation $\mathcal{A}_1$, the posterior we want to sample from is given by
\begin{equation}
\label{eqn:single-marg-post}
P(z \pbar \hat{\vct F}) \propto 
\mathcal{A}_1
\sum_k w^k 
c_1^k
\int \Theta(z, \vct{F})
\tilde{\mathcal{N}}(z, \vct{F} \pbar \vct{\eta}^{k \hat F}, \mtr{\Lambda}^{k \hat F})
\drv \vct{F} \,.
\end{equation}
However, the boundary prior makes this integral non-analytic and the resulting posterior is not a standard GMM, meaning that it cannot be sampled as described in section~\ref{sec:gmm-inference}.
Instead, we sample from the density given by }
\begin{equation}
\label{eqn:single-full-post}
P(z, \vct{F} \pbar \hat{\vct F}) \propto 
\mathcal{A}_1 \sum_k w^k 
c_1^k
\tilde{\mathcal{N}}(z, \vct{F} \pbar \vct{\eta}^{k \hat F}, \mtr{\Lambda}^{k \hat F}) \,.
\end{equation}
{This is the desired posterior from equation~\ref{eqn:single-marg-post} without the marginalisation over fluxes and where we have neglected the boundary prior term. This can then be corrected for by rejecting any sample that contains negative fluxes or redshift, leaving only the samples that obey the boundary prior. }
The marginalisation can then be done trivially by discarding the fluxes and considering only the redshift part of the remaining samples. Since equation~\ref{eqn:single-full-post} is simply a new Gaussian mixture model as before, sampling from this distribution is extremely computationally efficient, as detailed in section~\ref{sec:gmm-inference}.
As described above, {the inclusion of the boundary prior} is most important for the blended photo-z, though we include it here for completeness and consistency with the blended case later.

\subsection{Single-constituent evidence} \label{sec:gmm-single-evd}

One of the more computationally demanding aspects of the method of \cite{blendz} is the use of nested sampling in order to calculate the evidence. A significant advantage of the GMM method presented here is that this expensive integral can be evaluated much more quickly, an important feature for applying the method to future surveys.

The single-constituent evidence $\mathcal{E}^1$ is defined to be the integral of the unnormalised posterior over the full parameter space, i.e.,
\begin{equation}
\label{eqn:evidence-dfdz}
\begin{aligned}
\mathcal{E}^1
&= \int
\int P(\hat{\vct F} \pbar \vct{F}) \; P(z, \vct{F}) \drv \vct{F} 
\drv z  \,.
\end{aligned}
\end{equation}
As described above, by ignoring the boundary prior, the integral over fluxes can be performed analytically to give a new Gaussian mixture model. Inserting this result into the evidence integral, equation~\ref{eqn:evidence-dfdz} becomes
\begin{equation}
\begin{aligned}
\mathcal{E}^1
&= \sum_k w^k \;
% 
%\mathcal{N} (\vct{\mu}^k_{f} \pbar \hat{\vct F}, \mtr{\Sigma}^k_{\textrm{ff}} + \mtr{\Sigma}^{\hat F}) \;
c_1^k
\int
\mathcal{N}(z \pbar \vct{\mu}^{k \hat F}_\textrm{z}, \mtr{\Sigma}^{k \hat F}_{\textrm{zz}})  \drv z \,.
\end{aligned}
\end{equation}

Since the multivariate Gaussian density of each component is normalised to unity, the evidence is then given simply by the sum over the new mixture weights, i.e.,
\begin{equation}
\mathcal{E}^1 = \sum_k w^k 
% 
%\mathcal{N} (\vct{\mu}^k_{f} \pbar \hat{\vct F}, \mtr{\Sigma}^k_{\textrm{ff}} + \mtr{\Sigma}^{\hat F}) 
c_1^k
\equiv \sum_k w^{k \hat F}
\,.
\end{equation}
In this case, the evidence is analytic and therefore easy to compute. However, as above, computing these integrals analytically implicitly involves contributions from non-physical negative fluxes and redshifts. 

%While contributions from negative fluxes are heavily down-weighted by the likelihood, the same is only true of negative redshifts if the observed fluxes are well described by the single-component model. As described in section~\ref{sec:gmm-train}, the mixture prior allows negative redshifts, and if they are not down-weighted, they could contribute significantly to the evidence integral.

%In practice, negative redshifts are \textit{a posteriori} unlikely for sources of a single constituent galaxy that are well represented by the training set. If a source is very different from the training set, its inferred redshift might be expected to be very inaccurate anyway. However, a blended source that is badly fit by assuming a single constituent should be identified easily. Negative redshifts are therefore a problem here, as their contribution to the evidence integral makes inferring the source is unblended \textit{a-posteriori} more likely. We therefore need to adopt an additional boundary prior as described in section~\ref{sec:gmm-train}.

To combat this, we can numerically integrate the non-marginalised posterior of fluxes and redshifts including the boundary prior introduced in section~\ref{sec:gmm-train} and the accompanying normalisation from equation~\ref{eqn:norm1}, i.e., 
\begin{equation}
\mathcal{E}^1 =  
\int \!\!\! \int
\mathcal{A}_1
\sum_k w^k 
% \mathcal{N} (\vct{\mu}^k_{f} \pbar \hat{\vct F}, \mtr{\Sigma}^k_{\textrm{ff}} + \mtr{\Sigma}^{\hat F})  
c_1^k
\Theta(z, \vct{F})
\tilde{\mathcal{N}}(z, \vct{F} \pbar \vct{\eta}^{k \hat F}, \mtr{\Lambda}^{k \hat F}) 
\drv \vct{F} \drv z
\,.
\end{equation}
This integral can be evaluated {numerically} by using fluxes and redshifts sampled from the non-marginalised posterior with the boundary prior removed, given in equation~\ref{eqn:single-full-post}. This is another Gaussian mixture model, and thus these samples are computationally efficient to draw, as described in section~\ref{sec:gmm-inference}. In addition, the posterior samples drawn for inference {are also sampled from equation~\ref{eqn:single-full-post} and so can be reused here}, saving computation. 

Given a set of samples $\{z, \vct{F}\}$ from equation~\ref{eqn:single-full-post}, only a fraction $\mathcal{F}_1$ of these will contain no negative fluxes. Unlike equation~\ref{eqn:norm1}, however, this density is not normalised to unity, but rather
\begin{equation}
\begin{aligned}
\mathcal{V}_1 &\equiv
\mathcal{A}_1
\sum_k w^k 
c_1^k
\int \!\!\! \int \tilde{\mathcal{N}}(z, \vct{F} \pbar \vct{\eta}^{k \hat F}, \mtr{\Lambda}^{k \hat F}) 
\drv z \drv \vct{F} 
\\&= 
\mathcal{A}_1
\sum_k w^k 
c_1^k
\,.
\end{aligned}
\end{equation}
By using this to compute a Monte Carlo estimate of the integral, the evidence can therefore be estimated to be
\begin{equation}
\mathcal{E}^1 \approx  
\mathcal{V}_1 \mathcal{F}_1
= 
\mathcal{A}_1 \mathcal{F}_1
\sum_k w^k c_1^k
\equiv 
\mathcal{A}_1 \mathcal{F}_1
\sum_k w^{k \hat F} \,.
\end{equation}
%

%This would seem to require a set of samples from every component. However, a simple approximation can be made to reduce the computational cost. \note{As described in section~\ref{sec:gmm-inference}}, sampling the full posterior GMM can be accomplished by first sampling counts $n^k$ for the each component given its weight, and then sampling a set of fluxes and redshifts from the respective multivariate Gaussian $n^k$ times i.e., $\{ z_i, \vct{F}_i \pbar k, i=1\dots n^k \}$. However, many components have such low \textit{a posteriori} weight that they are not sampled at all, i.e., $n^k=0$. 

%A simple approximation for $\mathcal{F}^{k \hat F}$ is to calculate this fraction from the corresponding posterior samples $\{ z_i, \vct{F}_i \pbar k, i=1\dots n^k \}$, and assume that $\mathcal{F}^{k \hat F}=1$ when $n^k=0$. This approximation works because, even though the unsampled components contribute too highly to the evidence, their contribution (i.e., their \textit{a posteriori} weight $w^{k \hat F}$) must be small for them to have not been sampled. Similarly, the approximation of equation~\ref{eqn:approx-int-f} is most accurate for well sampled components which, by definition, are the components contributing the most to the evidence. 

%By making this approximation, the posterior samples drawn for the parameter inference step can be reused to calculate the evidence. \note{End up with some sources with no samples}

\subsection{Two-constituent posterior} \label{sec:gmm-blend-post}

We now extend the inference method to the case of a blended source consisting of two constituent galaxies by deriving the two-constituent posterior. 
Here, the parameters we wish to infer are the redshifts of each constituent $\{z \} = \{z_1, z_2 \}$, given the data vector of observed fluxes $\hat{\vct F}$.

As before, we start by marginalising over the latent flux vectors. As this is the two-constituent posterior, there are now two flux vectors to marginalise over, $\{ \vct{F} \} = \{ \vct{F}_1, \vct{F}_2 \}$, one for each galaxy. The posterior is therefore given by
\begin{equation}
\begin{aligned}
P(\{z\} \pbar \hat{\vct F}) 
&= \int P(\{z\}, \{\vct{F}\} \pbar \hat{\vct F}) \drv \{\vct{F}\} \,.
\end{aligned}
\end{equation}
Applying Bayes rule, this becomes
\begin{equation}
\label{eqn:post2-bayes}
\begin{aligned}
P(\{z\} \pbar \hat{\vct F}) 
&\propto \int P(\hat{\vct F} \pbar \{\vct{F}\}) \; P(\{z\}, \{\vct{F}\}) \drv \{\vct{F}\} \,,
\end{aligned}
\end{equation}
where $P(\{z\}, \{\vct{F}\})$ is the joint prior over flux and redshift for both constituents. This prior can be factorised to be written in terms of the individual constituent priors $ P(z, \vct{F})$, allowing the GMM to be inserted. However, as described in section~\ref{sec:blendz}, the parameters of each constituent are correlated. Thus, the joint prior can be written as
\begin{equation}
\label{eqn:joint-prior-propto}
\begin{aligned}
P(\{z\}, \{\vct{F}\})
&\propto   P(z_1, \vct{F}_1) P(z_2, \vct{F}_2) 
M(z_1, z_2) \,,
\end{aligned}
\end{equation}
where the blending-related correlations have been factored into a single term
\begin{equation}
M(z_1, z_2) = \pi(z_1, z_2) \; [1 + \xi(z_1, z_2)] \,.
\end{equation}
Here, $\xi(z_1, z_2)$ is the two-point galaxy correlation function, evaluated at the line-of-sight comoving distance between $z_2$ and $z_1$. {This correlation function is commonly modelled as a power law~\citep[e.g.,][]{xiDefinition}. However, we make no assumption of its form throughout this derivation, requiring only that it can be evaluated given a pair of redshifts. This correlation function was found to have little effect in \cite{blendz} so the results throughout assume $\xi(z_1, z_2)=0$. Nevertheless, we include it in the derivations here for completeness. The term} $\pi(z_1, z_2)$ represents the sorting condition, given by
\begin{equation}
\pi(z_1, z_2)= 
\begin{cases}
1 \quad \textrm{for } z_1 \leq z_2
\\
o \quad \textrm{otherwise.}
\end{cases}
\end{equation}
The need for these terms is briefly described in section~\ref{sec:blendz}; see section 2.3 of \cite{blendz} for more details. 

Any selection effects on the training set are already captured in the prior through the training step. {This assumes that the training set is sufficiently representative of the test set, though we note that this caveat applies to machine learning methods in general.} The selection effect term of \cite{blendz} simply acts to disfavour inferring fluxes such that the total flux is near the survey limit, as they are \textit{a priori} less likely to have been selected. Since the total flux is well constrained by observations, this term has little effect on parameter inferences. Instead, its use is motivated by making the magnitude prior proper. This is necessary for evaluating the marginal likelihood for model comparison. However, our GMM prior is proper by construction. As a result, we do not include the selection effect term here.

%It might be expected that constituent galaxies in a blended source will experience different selection effects to those that are unblended because they are selected on the total blended flux of the source, a quantity that they contribute only a fraction of. In this case, substantial differences in the flux-redshift distributions of individual constituents in a blended source can be reflected in the GMM by varying the mixture weights between single and unblended sources. 

As in section~\ref{sec:gmm-single-post}, the model selection requires the joint prior to be normalised. We do this by integrating the prior using Monte Carlo integration. To be able to draw samples from the prior efficiently, we insert the definitions of each term and combine into another Gaussian mixture that can be sampled as described in section~\ref{sec:gmm-inference}.
We also include the boundary prior described in section~\ref{sec:gmm-train} in each constituent prior to prevent contributions to the density from non-physical negative fluxes and redshifts. 
%While this is negligible in the single constituent case, the effect is more severe here. The only observable quantity in blended sources is the total flux, meaning that the likelihood term does not strongly constrain each flux vector independently. Without this, negative fluxes would be able to contribute strongly to the density.

Inserting the GMM, correlation and boundary prior terms into equation~\ref{eqn:joint-prior-propto}, the joint prior becomes
\begin{equation}
\label{eqn:joint-prior}
\begin{aligned}
P(\{z\}, \{\hat{\vct F}\})	
& \propto
M(z_1, z_2)
\Theta(z_1, \vct{F}_1) \Theta(z_2, \vct{F}_2)
\sum_k \sum_j w^k w^j 
\times \\ &\indent
\mathcal{N} (z_1, \vct{F}_1 \pbar \vct{\mu}^k, \mtr{\Sigma}^k)
\mathcal{N} (z_2, \vct{F}_2 \pbar \vct{\mu}^j, \mtr{\Sigma}^j) \,.
\end{aligned}
\end{equation}
We now follow an analogous method to that of section~\ref{sec:gmm-single-post} by combining the two multivariate Gaussians into a single density.
We start by defining a partitioned parameter vector that each density can be written in terms of. This is given by
\begin{equation}
	\label{eqn:blend-par-vect}
	\vct{\theta} = \left( \begin{matrix}
		z_1 \\
		\vct{F}_1 \\
		z_2 \\
		\vct{F}_2 \\
	\end{matrix}\right) \,.
\end{equation}
The product of the densities in equation~\ref{eqn:joint-prior} can then be written as a single Gaussian density in terms of this parameter vector
\begin{equation}
	\mathcal{N} (z_1, \vct{F}_1 \pbar \vct{\mu}^k, \mtr{\Sigma}^k)  \; 
	\mathcal{N} (z_2, \vct{F}_2 \pbar \vct{\mu}^j, \mtr{\Sigma}^j)  \; 
	=
	\mathcal{N} (\vct{\theta} \pbar \vct{\mu}^{kj}, \mtr{\Sigma}^{kj}) \,,
\end{equation}
where the new mean vector is given by
\begin{equation}
	\vct{\mu}^{kj} = 
	\left(
	\begin{matrix} 
		\vct{ \mu}^k \\
		\vct{ \mu}^j \\
	\end{matrix}
	\right)
	= 
	\left(
	\begin{matrix} 
		\mu^k_\textrm{z} \\
		\vct{ \mu}^k_\textrm{f} \\
		\mu^j_\textrm{z} \\
		\vct{ \mu}^j_\textrm{f} \\
	\end{matrix}
	\right)
	%\mathcal{N} (\vct{\theta} \pbar \vct{\mu}^{kj}, \mtr{\Sigma}^{kj})
\end{equation}
and the covariance matrix
\begin{equation}
	\label{eqn:sigma-kj}
	\mtr{\Sigma}^{kj} = 
	\left(
	\begin{matrix} 
		\mtr{\Sigma}^k &&  \mtr{0} \\
		\mtr{0} &&  \mtr{\Sigma}^j \\
	\end{matrix}
	\right)
	= 
	\left(
	\begin{matrix} 
		{\Sigma}^k_{\textrm{zz}} && \vct{\Sigma}^k_{\textrm{zf}} && {0} && \vct{0}\\
		\vct{\Sigma}^k_{\textrm{fz}} && \mtr{\Sigma}^k_{\textrm{ff}} && \vct{0} && \mtr{0}\\
		{0} && \vct{0} && {\Sigma}^j_{\textrm{zz}} && \vct{\Sigma}^j_{\textrm{zf}}\\
		\vct{0} && \mtr{0} && \vct{\Sigma}^j_{\textrm{fz}} && \mtr{\Sigma}^j_{\textrm{ff}} \\
	\end{matrix}
	\right)
	\,.
\end{equation}
This combination is trivial since we assume that all correlations between the two constituents have already been factored out into $M(\{z\}, \{\vct{F}\})$. As a result, the two constituent priors are independent and can be combined with the block diagonal covariance matrix defined in equation~\ref{eqn:sigma-kj}. The joint prior thus becomes
\begin{equation}
	\label{eqn:joint-prior-combined}
	\begin{aligned}
		P(\{z\}, \{\hat{\vct F}\})	
		& \propto
		M(z_1, z_2)
		\Theta(z_1, \vct{F}_1) \Theta(z_2, \vct{F}_2)
		\times \\ &\indent
		\sum_k \sum_j w^k w^j 
		\mathcal{N} (\vct{\theta} \pbar \vct{\mu}^{kj}, \mtr{\Sigma}^{kj})  \,,
	\end{aligned}
\end{equation}
i.e., a GMM multiplied  by several additional terms. The normalisation of this prior is then given by the integral
\begin{equation}
	\label{eqn:joint-prior-integral}
	\begin{aligned}
		\mathcal{A}_2
		& = \int \!\!\! \int \!\!\! \int \!\!\! \int
		M(z_1, z_2)
		\Theta(z_1, \vct{F}_1) \Theta(z_2, \vct{F}_2)
		\times \\ &\indent
		\sum_k \sum_j w^k w^j 
		\mathcal{N} (\vct{\theta} \pbar \vct{\mu}^{kj}, \mtr{\Sigma}^{kj})  
		\drv z_1 \drv z_2 \drv \hat{\vct F}_\alpha \drv \hat{\vct F}_\beta \,.
	\end{aligned}
\end{equation}
Analogously to equation ~\ref{eqn:norm1}, this can be evaluated using samples drawn from the Gaussian mixture, i.e.,
\begin{equation}
	\label{eqn:g-theta}
	\begin{aligned}
		\{
		z_1, z_2 , \vct{F}_1 , \vct{F}_2 
		\} \sim G(\vct{\theta}) =
		\sum_k \sum_j w^k w^j 
		\mathcal{N} (\vct{\theta} \pbar \vct{\mu}^{kj}, \mtr{\Sigma}^{kj})   \,.
	\end{aligned}
\end{equation}
Given $n_{\mathcal{A}}$ of these samples $\{z_1^i, z_2^i , \vct{F}_1^i , \vct{F}_2^i \pbar i = 1 \dots n_\mathcal{A} \}$, we can compute a Monte Carlo integration of $\mathcal{A}_2$ through importance sampling. Since $G(\vct{\theta})$ is normalised to unity, this integral is given by
\begin{equation}
\label{eqn:f2}
\mathcal{A}_2 = \sum_i
\frac{
[1 + \xi(z_1^i, z_2^i)]\; 
\pi(z_1^i, z_2^i) 
\Theta(z_1^i, \vct{F}_1^i) \Theta(z_2^i, \vct{F}_2^i)
}
{n_\mathcal{A}} \,.
\end{equation}
If the correlation function and sorting condition were ignored, this would simply be equal to the fraction of samples that obey the boundary prior, as in the definition of $\mathcal{A}_1$. Thus, the joint prior is given by
\begin{equation}
	\label{eqn:joint-prior-normed}
	\begin{aligned}
		P(\{z\}, \{\hat{\vct F}\})	
		& =
		\mathcal{A}_2
		M(z_1, z_2)
		\Theta(z_1, \vct{F}_1) \Theta(z_2, \vct{F}_2)
		\times \\ &\indent
		\sum_k \sum_j w^k w^j 
		\mathcal{N} (\vct{\theta} \pbar \vct{\mu}^{kj}, \mtr{\Sigma}^{kj}) \,.
	\end{aligned}
\end{equation}
This joint prior can then be inserted into equation~\ref{eqn:post2-bayes} alongside the definition of the likelihood to develop the posterior. As before, we assume that the likelihood is a multivariate Gaussian centred on the observed fluxes, though we now model the flux as the sum of the constituent fluxes, i.e.,  
\begin{equation}
P(\hat{\vct F} \pbar \vct{F}) = \mathcal{N}(\vct{F}_1 + \vct{F}_2 \pbar \hat{\vct F} , \mtr{\Sigma}^{\hat F} )
\,.
\end{equation}
Inserting this likelihood and the joint prior into equation~\ref{eqn:post2-bayes}, the posterior becomes
\begin{equation}
\begin{aligned}
P(\{z\} \pbar \hat{\vct F}) 
\propto & \mathcal{A}_2
\int \!\!\! \int
M(z_1, z_2) 
\Theta(z_1, \vct{F}_1) \Theta(z_2, \vct{F}_2)
\times \\ &
\sum_k \sum_j w^k w^j 
\mathcal{N}(\vct{F}_1 + \vct{F}_2 \pbar \hat{\vct F} , \mtr{\Sigma}^{\hat F} )
\times \\ &
\mathcal{N} (\vct{\theta} \pbar \vct{\mu}^{kj}, \mtr{\Sigma}^{kj}) 
%%%%%%%%%
\drv \vct{F}_1 \drv \vct{F}_2\,.
\end{aligned}
\end{equation}

To combine the prior term with the likelihood, we rewrite it in terms of natural parameters partitioned in the same way as equation~\ref{eqn:blend-par-vect}. These new parameters are given by
\begin{equation}
\vct{\eta}^{kj} = 
\left(
\begin{matrix} 
\vct{ \eta}^k \\
\vct{ \eta}^j \\
\end{matrix}
\right)
= 
\left(
\begin{matrix} 
\eta^k_\textrm{z} \\
\vct{ \eta}^k_\textrm{f} \\
\eta^j_\textrm{z} \\
\vct{ \eta}^j_\textrm{f} \\
\end{matrix}
\right)
\end{equation}
and 
\begin{equation}
\mtr{\Lambda}^{kj} 
= 
\left(
\begin{matrix} 
\mtr{\Lambda}^k &&  \mtr{0} \\
\mtr{0} &&  \mtr{\Lambda}^j \\
\end{matrix}
\right)
= 
\left(
\begin{matrix} 
{\Lambda}^k_{\textrm{zz}} && \vct{\Lambda}^k_{\textrm{zf}} && {0} && \vct{0}\\
\vct{\Lambda}^k_{\textrm{fz}} && \mtr{\Lambda}^k_{\textrm{ff}} && \vct{0} && \mtr{0}\\
{0} && \vct{0} && {\Lambda}^j_{\textrm{zz}} && \vct{\Lambda}^j_{\textrm{zf}}\\
\vct{0} && \mtr{0} && \vct{\Lambda}^j_{\textrm{fz}} && \mtr{\Lambda}^j_{\textrm{ff}} \\
\end{matrix}
\right)
\,.
\end{equation}
By also rewriting the likelihood in terms of the natural parameters 
$
\mtr{\Lambda}^{\hat F} \equiv \left(\mtr{\Sigma}^{\hat F}\right)^{-1}
$
and
$
\vct{\eta}^{\hat F} \equiv \mtr{\Lambda}^{\hat F}  \hat{\vct F}
$, 
the posterior becomes
\begin{equation}
	\begin{aligned}
		P(\{z\} \pbar \hat{\vct F}) 
		& \propto \int \!\!\! \int
		\mathcal{A}_2
		M(z_1, z_2)
		\Theta(z_1, \vct{F}_1) \Theta(z_2, \vct{F}_2)
		\times \\ &\indent
		\sum_k \sum_j w^k w^j 
		\tilde{\mathcal{N}}(\vct{F}_1 + \vct{F}_2 \pbar \vct{\eta}^{\hat F}, \mtr{\Lambda}^{\hat F})
		\times \\ &\indent
		\tilde{\mathcal{N}} (\vct{\theta} \pbar \vct{\eta}^{kj}, \mtr{\Lambda}^{kj}) 
		%%%%%%%%%
		\drv \vct{F}_1 \drv \vct{F}_2\,.
	\end{aligned}
\end{equation}
The two remaining densities can now be combined into a single term given by
\begin{equation}
\label{eqn:combine-joint-prior-like}
\tilde{\mathcal{N}}(\vct{F}_1 + \vct{F}_2 \pbar \vct{\eta}^{\hat F}, \mtr{\Lambda}^{\hat F})  \; 
\tilde{\mathcal{N}} (\vct{\theta} \pbar \vct{\eta}^{kj}, \mtr{\Lambda}^{kj})
\propto \tilde{\mathcal{N}} (\vct{\theta} \pbar \vct{\eta}^{k \hat F}, \mtr{\Lambda}^{k \hat F}) \,,
\end{equation}
where the combined parameters are given by
\begin{equation}
\vct{\eta}^{k \hat F} = 
\left(
\begin{matrix} 
\eta_\textrm{z}^{k} \\
\vct{\eta}_\textrm{f}^{k} + \vct{\eta}^{\hat F} \\
\eta_\textrm{z}^{j} \\
\vct{\eta}_\textrm{f}^{j} + \vct{\eta}^{\hat F}\\
\end{matrix}
\right)
\end{equation}
and
\begin{equation}
\mtr{\Lambda}^{k \hat F} = 
\left(
\begin{matrix} 
{\Lambda}^k_{\textrm{zz}} && \vct{\Lambda}^k_{\textrm{zf}} && {0} && \vct{0}\\
\vct{\Lambda}^k_{\textrm{fz}} && \mtr{\Lambda}^k_{\textrm{ff}} + \mtr{\Lambda}^{\hat F} && \vct{0} && \mtr{\Lambda}^{\hat F}\\
{0} && \vct{0} && {\Lambda}^j_{\textrm{zz}} && \vct{\Lambda}^j_{\textrm{zf}}\\
\vct{0} && \mtr{\Lambda}^{\hat F} && \vct{\Lambda}^j_{\textrm{fz}} && \mtr{\Lambda}^j_{\textrm{ff}} + \mtr{\Lambda}^{\hat F} \\
\end{matrix}
\right) \,.
\end{equation}
As before, the constant of proportionality $c_2^{kj}$ in equation~\ref{eqn:combine-joint-prior-like} can also be written in terms of another multivariate Gaussian density
\begin{equation}
\begin{aligned}
c_2^{kj}
&= 
\mathcal{N}
\left(
\mu^{k}_\textrm{f}+\mu^{j}_\textrm{f} 
\pbar
\hat{\vct F}, 
\left[
\Sigma^{\hat F} 
+ \Sigma^{k}_{\textrm{ff}}
+ \Sigma^{j}_{\textrm{ff}}
\right]
\right) \,.
\end{aligned}
\end{equation}
The posterior is thus given by
\begin{equation}
	\label{eqn:blend-post-marg}
	\begin{aligned}
		P(\{z\} \pbar \hat{\vct F}) 
		& \propto 
		\int \!\!\! \int
		\mathcal{A}_2
		M(z_1, z_2) 
		\Theta(z_1, \vct{F}_1) \Theta(z_2, \vct{F}_2)
		\times \\ &\quad
		\sum_k \sum_j w^k w^j c_2^{kj}
		\mathcal{N} (\vct{\theta} \pbar \vct{\eta}^{k \hat F}, \mtr{\Lambda}^{k \hat F}) 
		%%%%%%%%%
		\drv \vct{F}_1 \drv \vct{F}_2\,.
	\end{aligned}
\end{equation}

As in the single constituent case, it would be possible to do this integral analytically by ignoring the boundary prior $\Theta(z, \vct{F})$. Converting back to the standard parametrisation, the final posterior would then be given by
\begin{equation}
	\begin{aligned}
		P(\{z\} \pbar \hat{\vct F}) 
		& \propto 
		\mathcal{A}_2
		M(z_1, z_2) 
		\sum_k \sum_j w^k w^j c_2^{kj}
		\mathcal{N} (z_1, z_2 \pbar \vct{\mu}^{k \hat F}_\textrm{z}, \mtr{\Sigma}^{k \hat F}_{\textrm{zz}}) \,.
	\end{aligned}
\end{equation}

%However, as described in section~\ref{sec:gmm-single-post}, contributions from negative fluxes and redshifts can be substantial in the blended case. As a result, ignoring the boundary prior here is not the safe assumption it was in the single-constituent case. It is therefore important to compute this integral while including the boundary prior terms.
%
%Since this 
With the boundary prior, the integral is no longer analytically tractable. As a result, we take the same approach as in the single constituent case and sample from the full, non-marginalised posterior. An additional complication here are the extra correlations factored into $M(z_1, z_2)$. As a result of this term, the posterior is no longer a Gaussian mixture and therefore does not permit the efficient sampling scheme described in section~\ref{sec:gmm-inference}.

Instead, we can sample from the full posterior distribution ignoring the contribution of both the the boundary prior and the correlations, modifying the samples \textit{post hoc} by rejection and reweighting to correct for these respectively. This set of samples is thus drawn from the simplified posterior $H(\vct{\theta})$,  given by
\begin{equation}
	\label{eqn:h-theta}
	\begin{aligned}
	\{z_1, z_2 , \vct{F}_1 , \vct{F}_2 
		\} \sim H(\vct{\theta})
		\propto & \mathcal{A}_2
		\sum_k \sum_j w^k w^j c_2^{kj}
		\times \\
		&\mathcal{N} (\vct{\theta} \pbar \vct{\eta}^{k \hat F}, \mtr{\Lambda}^{k \hat F}) \,.
	\end{aligned}
\end{equation}
This simplified posterior is now a standard GMM, and can therefore be efficiently sampled as described in section~\ref{sec:gmm-inference}. 
The neglected terms can now be corrected for separately.

Firstly, the boundary priors can be included by rejecting samples where the flux or the redshift is negative, as in section~\ref{sec:gmm-single-post}. The sorting condition could also be included by simply rejecting samples where it was not respected. However, this is unnecessarily wasteful of computation. Note that mixture component-$jk$ is identical to component-$kj$ under exchange of constituents. Every component is matched with a pair in this way. As a result, the posterior is exactly symmetric, meaning that samples with misordered redshifts can be corrected by simply swapping the order of their constituents. 

%This exchangeability is a result of the priors of each constituent being the same. \note{REMOVED SEC 5} Section~??? extends the model so that this is no longer the case by allowing each constituent its own weight vector. Reordering the samples would therefore not work in this case, and would instead require rejections.

The redshift correlation function can be corrected for using importance sampling by associating each sample with a weight $[1 + \xi(z_1, z_2)]$. All inferences done with these samples would then need to account for these weights. The risk with this importance sampling approach is that regions of parameter space where the correlation function is large could be poorly sampled when using the modified posterior. The effect of the correlation function would then be under-represented. However, \cite{blendz} found that including the redshift correlation function when sampling the posterior had little effect on inferences. As a result, we expect any errors from the use of importance sampling here to be negligible.

Given a set of corrected samples of redshift and flux, the marginalisation can then be done in the same way as in section~\ref{sec:gmm-single-post}, by discarding the flux parts of the samples. The distribution of the remaining redshift samples will then be proportional to the marginalised posterior defined in equation~\ref{eqn:blend-post-marg}, as desired.

\subsection{Two-constituent evidence} \label{sec:gmm-blend-evd}

The two-constituent evidence $\mathcal{E}^2$ is defined as the integral of the blended posterior over both sets of fluxes and redshifts, i.e., 
\begin{equation}
	\label{eqn:evd2_define}
	\begin{aligned}
		\mathcal{E}^2 
		= \int \!\!\! \int P(\hat{\vct F} \pbar \{\vct{F}\}) \; P(\{z\}, \{\vct{F}\}) \drv \{z\} \drv \{\vct{F}\} \,.
	\end{aligned}
\end{equation}
Inserting the definitions of each term from the full posterior given in equation~\ref{eqn:blend-post-marg}, this expression becomes
\begin{equation}
	\label{eqn:evd2-insert}
	\begin{aligned}
		\mathcal{E}^2
		& = 
		\mathcal{A}_2
		\int \!\!\! \int \!\!\! \int \!\!\! \int
		M(z_1, z_2) 
		\Theta(z_1, \vct{F}_1) \Theta(z_2, \vct{F}_2)
		\times \\ &\quad
		\sum_k \sum_j w^k w^j c_2^{kj}
		\mathcal{N} (\vct{\theta} \pbar \vct{\eta}^{k \hat F}, \mtr{\Lambda}^{k \hat F}) 
		%%%%%%%%%
		\drv z_1 \drv z_2
		\drv \vct{F}_1 \drv \vct{F}_2
		\,.
	\end{aligned}
\end{equation}
%
%As in the previous section, the correlations and boundary priors mean that this integral cannot be calculated analytically. We therefore adopt the same approach as that of section~\ref{sec:gmm-single-evd} for the single-constituent evidence and 
As before, we 
evaluate this integral numerically using Monte Carlo integration. 
To do this, we can reuse the samples drawn for the blended posterior inference from $H(\vct{\theta})$ defined in equation~\ref{eqn:h-theta}. Given a set of $n_2$ of these samples $\{z_1^i, z_2^i , \vct{F}_1^i , \vct{F}_2^i \pbar i = 1 \dots n_2 \}$, we can define the weighted fraction 
\begin{equation}
	\label{eqn:evd2-weighted-fraction}
	\mathcal{F}_2 = \sum_i
	\frac{
		[1 + \xi(z_1^i, z_2^i)]\; 
		%\Lambda(z_1^i, z_2^i) 
		\pi(z_1^i, z_2^i) 
		\Theta(z_1^i, \vct{F}_1^i) \Theta(z_2^i, \vct{F}_2^i)
	}
	{n_2} \,.
\end{equation}
This is analogous to $\mathcal{F}_1$, the fraction of samples drawn from the non-marginalised single-constituent posterior defined in equation~\ref{eqn:single-full-post} that obey the boundary prior, but with the additional blending-related correlations. The simplified posterior $H(\vct{\theta})$ is not normalised to unity. However, the normalisation constant $\mathcal{V}_2$ is given by the integral over the full support of the distribution, giving
\begin{equation}
	\label{eqn:v2}
	\begin{aligned}
		\mathcal{V}_2 &\equiv
		\int \mathcal{A}_2
		\sum_k \sum_j w^k w^j c_2^{kj}
		\mathcal{N} (\vct{\theta} \pbar \vct{\eta}^{k \hat F}, \mtr{\Lambda}^{k \hat F})
		\drv \vct{\theta}
		\\ &= 
		\mathcal{A}_2
		\sum_k \sum_j w^k w^j c_2^{kj}
		 \,.
	\end{aligned}
\end{equation}
Thus, the two-constituent evidence can be estimated by importance sampling to be
\begin{equation}
	\label{eqn:evd2}
	\begin{aligned}
		\mathcal{E}^2 \approx
		\mathcal{V}_2 \mathcal{F}_2
		= 
		\mathcal{A}_2 \mathcal{F}_2
		\sum_k \sum_j w^k w^j c_2^{kj}
		\,.
	\end{aligned}
\end{equation}

\section{Tests on simulated sources} \label{sec:sim-results}

In order to test our method, we construct a two sets of simulated observations to train our model and compare predictions against. These two sets correspond to an LSST-like optical survey~\citep{lsstSummary}, and the same survey with additional Euclid-like infrared observations~\citep{euclidSummary}. The complementarity of LSST and Euclid has been investigated previously~\citep[e.g.,][]{synergy}; additional filter bands will help to break colour-redshift degeneracies and therefore enable more accurate photometric redshifts.

Simulated observations are generated by redshifting a template, integrating over the relevant filter response curves, scaling the results to a given $i$-band magnitude, adding observational noise and imposing selection criteria. We use the set of templates assembled by \cite{coeTemplates} containing eight templates. 
%From \cite{colemanTemplates}, there are two templates representing late type galaxies, and one template each for early and irregular galaxies. The set also contains four starburst templates, two from \cite{kinneyTemplates}, and the rest added by \cite{coeTemplates}. 
This is the default template set in the commonly used BPZ~\citep{bpz} photometric redshift software.

We randomly sample true redshift, magnitude and template parameters for each source from a prior using \texttt{emcee}~\citep{emcee}. The single-constituent joint redshift-magnitude-template prior is defined as follows. First we factorise into separate prior terms, i.e., 
\begin{equation}
P(z, m, t) = P(z \pbar m) P(t \pbar m) P(m) \,,
\end{equation}
where $t$ is an integer labelling each template and the redshift prior is assumed to be independent of template. The redshift and magnitude priors are then given by the LSST predictions in \cite{lsstScienceBook}. The redshift prior, based on simulated high-redshift galaxy populations~\citep{highZSims} is given by
\begin{equation}
P(z \pbar m) = \frac{1}{2 z_0(m)} \left( \frac{z}{z_0(m)} \right)^2 \exp \left( \frac{-z}{z_0(m)} \right) \,,
\end{equation}
where
\begin{equation}
z_0 (m) = 0.0417 m - 0.744 \,,
\end{equation}
and $m$ refers to $i$-band magnitude. The corresponding $i$-band magnitude prior, fitted to data from the Canada-France-Hawaii Telescope Legacy Survey~\cite[CFHTLS;][]{cfhtls}, is then given by
\begin{equation}
P(m) \propto 10^{0.31(m - 25)} \,.
\end{equation}
We also use the template prior from \cite{bpz}, given by
\begin{equation}
P \Big(t \,\Big|\,  m \Big) = f_\textrm{t} \exp \left(- k_\textrm{t} [m - m_0 ] \right) \,,
\end{equation}
where we set $m_0 = 20$ and the parameters $f_\textrm{t}$ and $k_\textrm{t}$, each dependent on the template type, are set to the values given in \cite{bpz}.

Once the redshift, magnitude and template are sampled from this joint prior, the intrinsic fluxes are simulated by redshifting the template and integrating over filter response curves. For the optical survey, we use the six LSST filters $u,g,r,i,z,Y$~\citep{lsstScienceBook}. We use the three Euclid filters $Y, J, H$~\citep{euclidDesign} as additional infrared bands, giving a total of nine bands for the combined surveys.

Finally, we add magnitude-dependent observational noise to each band. For the optical bands, this is given by the predicted LSST noise model~\citep{lsstScienceBook}. The $5 \sigma$ depth of point sources in the Euclid $Y$, $J$ and $H$ bands is $24 \textrm{mag}$~\citep{euclidSummary}, the same depth as point sources in the LSST $i$-band~\citep{lsstScienceBook}. We therefore approximate the observational noise in the $Y$, $J$ and $H$ bands by assuming that their signal-to-noise is equal to that of the $i$-band.

In order to simulate the flux of blended sources, we add the intrinsic fluxes of two simulated sources and add observational noise corresponding to the total blended flux. The two-constituent prior also needs to account for the blended-related terms described above. The redshift prior includes the sorting condition $\pi(z_1, z_2)$, though we assume no clustering, i.e., $\xi(z_1, z_2)=0$, as it has a negligible effect at large separations when $z_1 \not\approx z_2$. We also impose a prior on the faintest $i$-band magnitude of either constituent such that it must be brighter than a 5$\sigma$ detection. A cut like this is necessary since it only makes sense to consider a source blended when each constituent is sufficiently bright. If a constituent is too faint, it should instead be considered to be a contributor to the background flux, rather than that of the source itself.

Finally, we select sources by imposing an $i$-band magnitude cut of $m_i < 25$. This corresponds to the LSST gold sample~\citep{lsstScienceBook}, a population of $\approx 4 \times 10^9$ high signal-to-noise galaxies. For each of the two sets of simulated sources, we randomly select $10000$ single-constituent sources to act as a training set, a further $10000$ single-constituent sources for the unblended test set, and $10000$ two-constituent sources for the blended test set.

Given the unblended training set, we use the procedure described in section~\ref{sec:gmm-cross-val} to set the number of mixture components $N$. Using $3$-fold cross-validation, we test from $N=5$ to $N=100$ in multiples of $5$, measuring the RMS scatter $\sigma_\textrm{RMS}$ defined in equation~\ref{eqn:sig-rms} at each iteration. In order to evaluate this, we must define a way to calculate a point estimate $z_\textrm{p}$ from a set of $n_2$ samples $\{z_{\textrm{p}, i} \pbar i=1 \dots n_2\}$ drawn from the posterior defined in section~\ref{sec:gmm-single-post}. We therefore define this point estimate to be the mean of these samples, as this is equivalent to a Monte Carlo estimate of the expectation value of the redshift, i.e.,
\begin{equation}
\label{eqn:avg-redshift}
z_\textrm{p} \equiv \frac{1}{n_2} \sum_{i=1} z_{\textrm{p}, i} \approx \int  P(z \pbar \hat{\vct F}) \, z \drv z \,.
\end{equation}
 
\begin{figure}
	\includegraphics[width=\columnwidth]{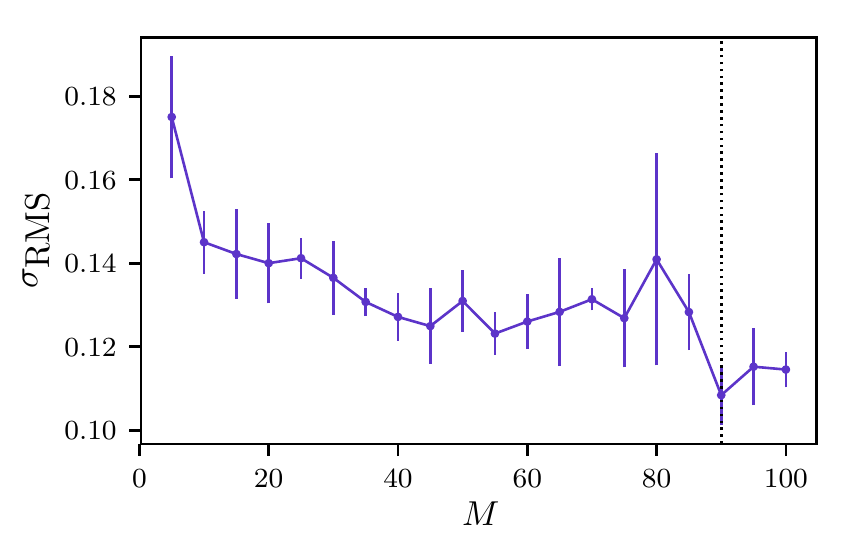}
	\caption{Results of the cross-validation for the LSST-like simulated data. The points show the RMS scatter averaged over the three folds, while the error bars show the error on the mean. We choose the number of components to be $N=90$, minimising the average RMS scatter as indicated by the dotted black line.}
	\label{fig:lsst-crossVal}
\end{figure}

The results of this cross-validation are shown in Fig.~\ref{fig:lsst-crossVal}. We find the average RMS scatter across all folds $\overline{ \sigma_\textrm{RMS}}$ to be minimised when $N=90$ with $\overline{ \sigma_\textrm{RMS}} = 0.108$. We therefore use a mixture comprised of 90 components to fit the entire training set for use throughout.

\begin{figure}
	\includegraphics[width=\columnwidth]{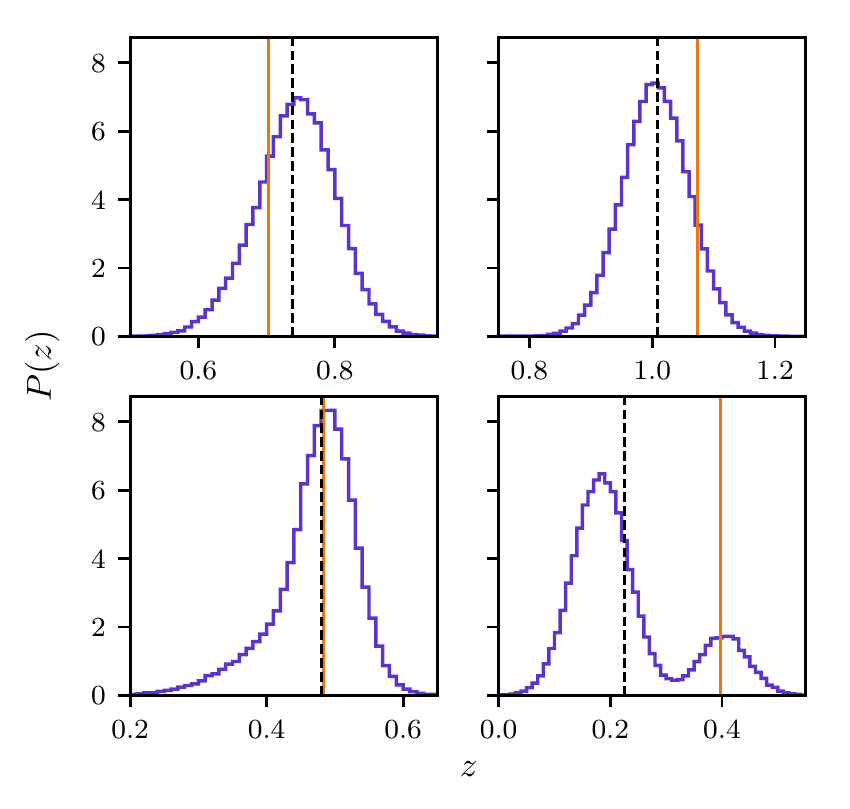}
	\caption{Plot showing four examples of single-constituent posteriors sampled using our method on the unblended LSST-like data. The black dashed lines indicate the sample means we use to define the point estimates $z_\textrm{p}$. The true redshifts are indicated by the {solid} orange lines.}
	\label{fig:lsst-single-post}
\end{figure}

Examples of one-constituent posteriors inferred using samples from the distribution defined in section~\ref{sec:gmm-single-post} and conditioned on the LSST-like data are shown in Fig.~\ref{fig:lsst-single-post}. The four panels in this figure show the variety of shapes of posteriors that can result from photometric redshifts and can be represented by the GMMs presented here. 

The top two panels of Fig.~\ref{fig:lsst-single-post} shows examples of well constrained, accurate posteriors; their shapes are symmetric and close to that of a single Gaussian. However, the posterior shown in the bottom left panel is left-skewed. This long-tailed posterior is a common occurrence in the results of photometric redshift inference. Despite being very non-Gaussian, it can be represented by a mixture of components. Finally, the bottom right panel shows an example of a bimodal posterior that can be easily represented by a mixture of well separated components. While the true redshift is contained well within the lower peak of this posterior, the bimodality has pulled the mean redshift to between the two peaks. As a result, the point estimate is inaccurate, despite the true redshift lying at a point of significantly non-zero posterior density. This demonstrates the loss of information resulting from the compression of a full posterior distribution to a single point estimate.

\begin{figure*}
	\includegraphics[width=0.32\textwidth]{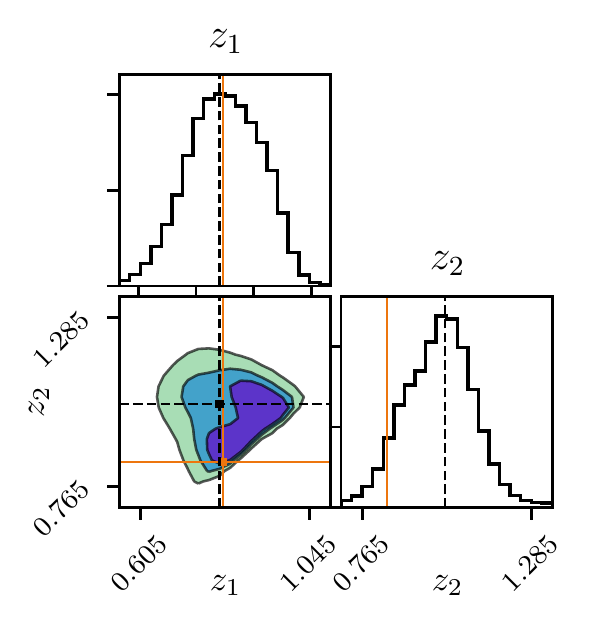}
	\includegraphics[width=0.32\textwidth]{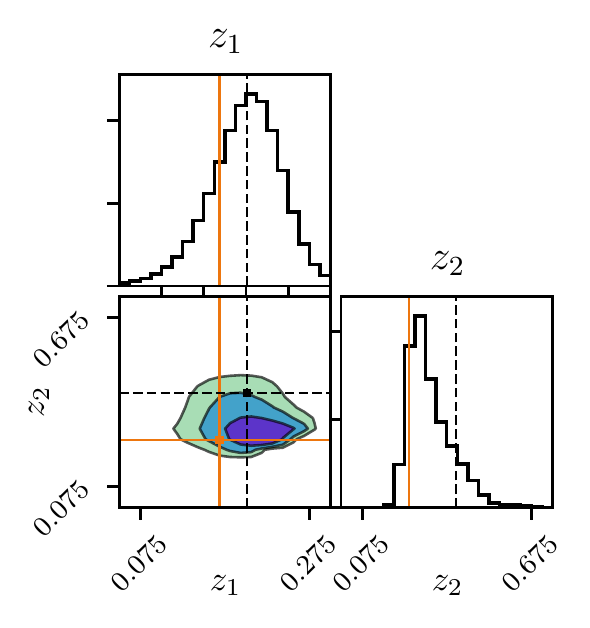}
	\includegraphics[width=0.32\textwidth]{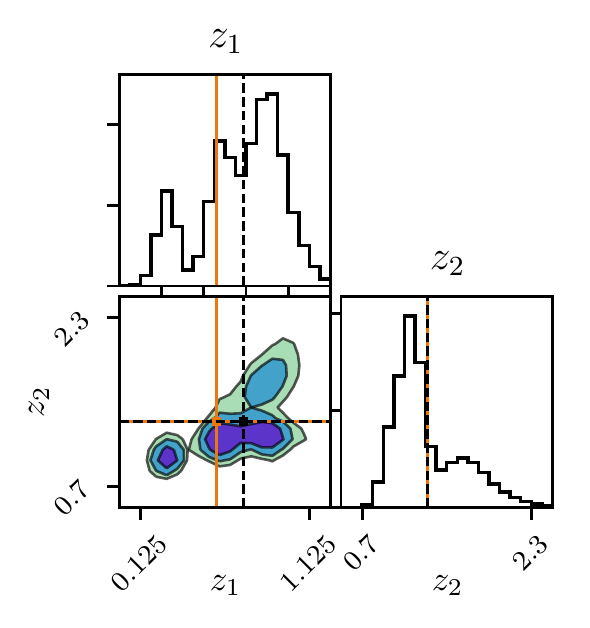}
	\caption{Plot showing three examples of two-constituent posteriors sampled using the GMM on the blended LSST-like data. The black dashed lines indicate the sample means we use to define the point estimates $z_\textrm{p}$. The true redshifts of each constituent are indicated by the orange lines.}
	\label{fig:lsst-blend-post}
\end{figure*}

Examples of two-constituent posteriors inferred using samples from the distribution defined in section~\ref{sec:gmm-blend-post} are shown in Fig.~\ref{fig:lsst-blend-post}. These samples are also drawn from posteriors conditioned on the LSST-like data.

The left panel of Fig.~\ref{fig:lsst-blend-post} shows a well constrained posterior. One edge of the joint distribution lies along the $z_1 = z_2$ line. As a result, the effect of the sorting condition $\pi(z_1, z_2)$ can be seen clearly, sharply cutting the joint distribution. The centre panel shows a joint posterior that results in highly skewed marginal distributions. As before, the long tail of the $z_2$ marginal distribution pulls the mean redshift away from the peak. This demonstrates that, since point estimates are inevitably less informative than the full posterior distribution, the choice of how these point estimates are defined can significantly alter their accuracy. In this case, the accuracy of the point estimate would be increased by choosing $z_2$ to be the redshift where the posterior peaks, i.e., the maximum \textit{a posteriori} (MAP) value. However, we found that MAP point estimates were less accurate over the whole sample on average. Finally, the right panel of Fig.~\ref{fig:lsst-blend-post} shows an example of a highly multimodal posterior that can arise in the two-constituent case.

\begin{figure*}
	\includegraphics[width=\textwidth]{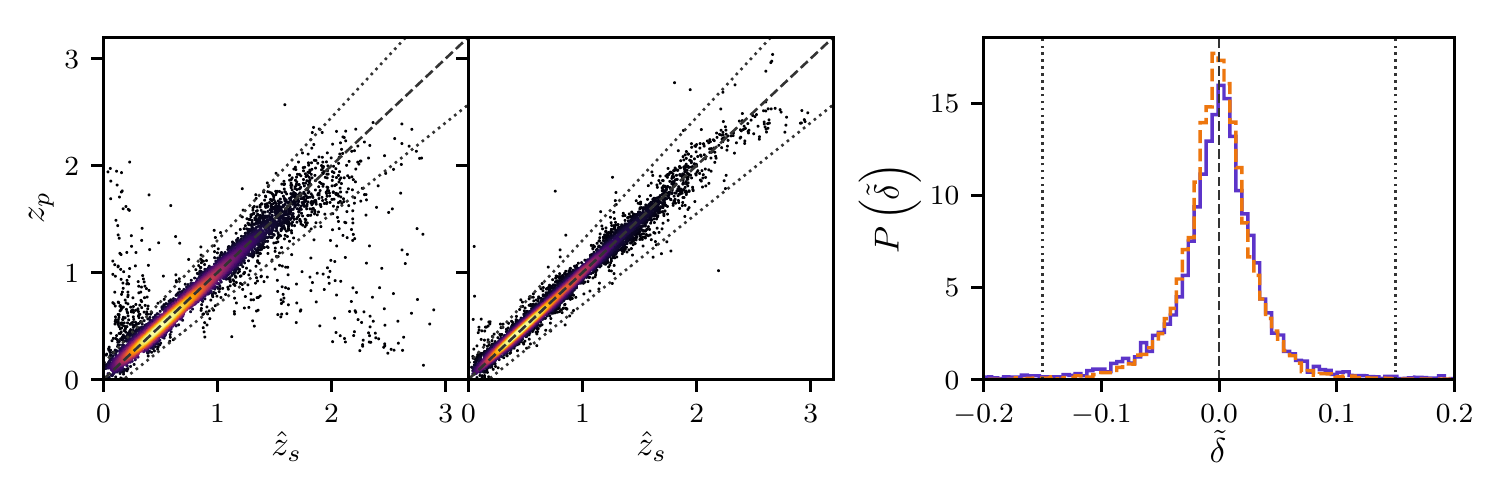}
	\caption{Plot showing the point-estimate results obtained from the GMM on the unblended simulated data. The left and right scatter plots show the point estimate results for the LSST-like and the combined LSST-Euclid-like surveys respectively. These plots show the benefit of additional bands and increased wavelength coverage from near-infrared data in reducing outliers. The dashed line denotes $z_\textrm{p} = \hat z_\textrm{s}$, and the dotted lines indicate our outlier definition where $|z_\textrm{p} - \hat z_\textrm{s}| \geq 0.15 (1 + \hat z_\textrm{s})$. Points are coloured according to their density on the scatter plots to illustrate overplotting. The right panel shows the distribution of the normalised error {$\tilde{\delta}$, defined in equation~\ref{eqn:norm-error}}. The solid purple line shows the results for the LSST-like survey, while the orange dashed line shows the results for the combined LSST-Euclid-like survey. The black dashed and dotted lines are defined as in the scatter plots.}
	\label{fig:lsst-euclid-single-scatter}
\end{figure*}

While less informative than the full posterior distributions, point estimates are still a common product of photometric redshift inference. A plot of these point estimates, defined as the mean of samples drawn from the posterior, against the true redshift for single-constituent data from the two simulated surveys is shown in Fig.~\ref{fig:lsst-euclid-single-scatter}.

This figure shows that the method performs well in the single-constituent case, i.e., on the standard photometric redshift inference problem. The vast majority of sources have their redshifts recovered accurately; this can be seen by the significant density of points around the $z_\textrm{p} = \hat z_\textrm{s}$ line, demonstrated in the plot by the colour of the points. Comparing the panels for the two simulations, the most significant difference is in the number of outliers, which is reduced in the simulations with additional infrared data. This can also be seen in the third panel, a histogram of the reduced error $\tilde{\delta}$ defined in equation~\ref{eqn:norm-error}. When zoomed around the majority of values at small errors, the difference between the histograms for the two sets of simulations is negligible.

This reduction of outliers is expected, as the additional filters can help to lift the colour-redshift degeneracies discussed in section~\ref{sec:gmm}. We define outliers to be sources where $|z_\textrm{p} - \hat z_\textrm{s}| \geq 0.15 (1 + \hat z_\textrm{s})$. This boundary is shown as a dotted line in Fig.~\ref{fig:lsst-euclid-single-scatter}.

In order to quantify the accuracy of these point estimates, we can use several metrics. Firstly, we use the RMS scatter defined in equation~\ref{eqn:sig-rms}. We find this scatter to be $\sigma_{\textrm{RMS}}=0.105$ for the LSST-like simulations, and $\sigma_{\textrm{RMS}}=0.038$ for the simulations with additional infrared data. While this difference is significant, it is primarily driven by the reduction of outliers by the infrared data.

In the LSST-like survey, $1.82 \%$ of sources are outliers. This is reduced to $0.10\%$ in the combined LSST-Euclid-like simulations. These outliers have significant errors by definition, are therefore can have a significant effect on the measured RMS scatter. In order to identify these outliers as the most significant driver of the difference in accuracy between the two sets of simulations, we measure the RMS scatter while neglecting these sources, as in the photometric redshift accuracy tests of ~\cite{photozAccuracy}. When this is done, the RMS of the LSST-like simulations drops to $\sigma_{\textrm{RMS}}=0.036$, while the scatter of the simulations with additional Euclid-like data becomes $\sigma_{\textrm{RMS}}=0.031$. Since these values are now far closer and the latter change was less dramatic, we conclude that the biggest benefit afforded by the additional bands is the reduction of outliers.

\begin{figure*}
	\includegraphics[width=\textwidth]{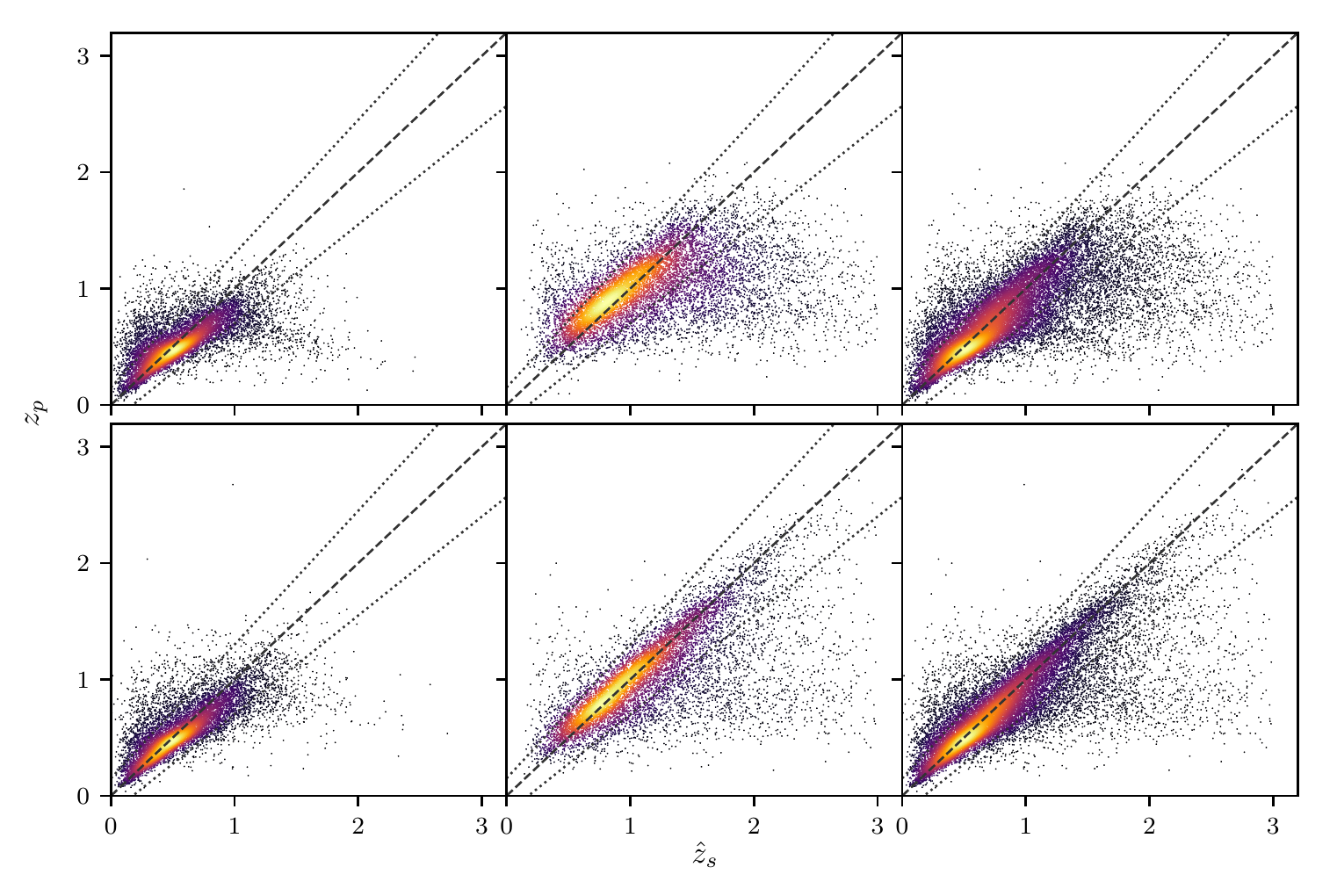}
	\caption{Plot showing the point-estimate results obtained from the GMM on the blended simulated data. The top row shows the results for the LSST-like survey, and the bottom row shows results for the combined LSST-Euclid-like survey. The left plots show $z_{\textrm{p}, 1}$, the point estimate of the redshift for the lower-redshift constituent in each blended source. The centre plots show $z_{\textrm{p}, 2}$, corresponding to the higher-redshift constituent in each blended source. The right plots combine both $z_{\textrm{p}, 1}$ and  $z_{\textrm{p}, 2}$. The dashed lines denotes $z_\textrm{p} = \hat z_\textrm{s}$, and the dotted lines indicate our outlier definition where $|z_\textrm{p} - \hat z_\textrm{s}| \geq 0.15 (1 + \hat z_\textrm{s})$. Points are coloured according to their density on the scatter plots to illustrate overplotting.}
	\label{fig:lsst-euclid-blend-scatter}
\end{figure*}

We also evaluate the same metrics on point estimates of the redshifts of the blended simulated data. These point estimates are defined to be the mean of posterior samples, as in the single-constituent case. A plot of these point estimates for each set of simulated data is shown in Fig.~\ref{fig:lsst-euclid-blend-scatter}.

The blended redshift inference is a more challenging problem than standard photometric redshifts of unblended sources. However, while the scatter plots in Fig.~\ref{fig:lsst-euclid-blend-scatter} are noisier than the single-constituent plots in Fig.~\ref{fig:lsst-euclid-single-scatter}, many redshifts are still recovered accurately. This can be seen in the high density of points around $z_\textrm{p} = \hat z_\textrm{s}$, again demonstrated by their colour. This increase in noise over the single-constituent case is expected, as the same number of data-points per source are used here to constrain twice the number of parameters.

As in the single-constituent case, the addition of additional bands in the infrared reduces both the RMS scatter and the number of outliers. For the LSST-like survey, we find the scatter to be $\sigma_{\textrm{RMS}}=0.171$, while the combined LSST-Euclid-like survey has a scatter of $\sigma_{\textrm{RMS}}=0.145$. The outlier rate of the former survey is found to be $17.5\%$, while that of the latter is reduced to $12.4\%$.

As discussed in section~\ref{sec:gmm}, an important part of the results of photometric redshift inference are PDFs. Unlike simple point estimates, PDFs represent the full statistical knowledge of the redshift being inferred and are essential for rigorously propagating uncertainties. It is therefore also important that the quality of the resulting PDFs are assessed. 

A conceptual problem with assessing the quality of PDFs is that there is no true PDF that they can be compared against. This is in contrast to point estimates where the spectroscopic redshift provides a known ground truth against which to compare. Instead, \cite{qqplot} introduce a frequentist method to test the widths of PDFs that relies on credible intervals (CIs).

The definition of CIs follows directly from that of posterior PDFs. For a given posterior $P(\theta \pbar d)$ that is correctly normalised, the conditional probability that the parameter $\theta$ will lie within an interval $[\theta_\textrm{low}, \theta_\textrm{high}]$ is given by the integral of the posterior over that interval, i.e.,
\begin{equation}
\label{eqn:define-ci}
P(\theta_\textrm{low} \leq \theta \leq \theta_\textrm{high} \pbar d) = \int_{\theta_\textrm{low}}^{\theta_\textrm{high}} P(\theta \pbar d) \drv \theta \,.
\end{equation}
The CI corresponding to a particular percentage is then defined to be the interval over which equation~\ref{eqn:define-ci} equals this percentage. In general, this interval will not be unique, since the integral over many different intervals can be the same. For this reason, the credible interval is often defined to be the highest posterior density (HPD) interval, the interval covering the shortest length in parameter space for a given integral. In general, this region does not need to be contiguous; the HPD region of multimodal posteriors will instead be made up of several subintervals.

A conceptually simple way to define this HPD region is to consider a horizontal line spanning the entirety of parameter space, drawn on a plot of the PDF. As this line is moved downwards, it will begin to intersect the PDF. The regions between these intersections can then be integrated to give an area. The intervals contained within these intersections are the HPD region corresponding to this area. Since this area will monotonically increase as the line is moved downwards, this provides a way to define the HPD region for a given percentage CI.

An intuitive interpretation of these intervals is that, given many repetitions of the experiment and the subsequent construction of many such intervals of area $\alpha$, the true parameter would be contained within a fraction $\alpha$ of these intervals. This notion is the interpretation of frequentist confidence intervals as coverage probabilities. However, while this interpretation is intuitive, it is not guaranteed by a Bayesian analysis. Instead, posteriors where this coverage probability property holds are said to be \textit{calibrated}, and several methods having been proposed to calibrate posteriors \citep[e.g.][]{bvmCalibrate, calibrateCosmo}.

The method introduced in \cite{qqplot} tests whether the posteriors resulting from a photometric redshift method are calibrated. If they are, we should expect that $50\%$ of sources have their true redshift within their $50\%$ CI. The equivalent statement can be made for all levels of CI, generalising this to a continuous test. The method may therefore give an indication of the performance of the method, and such a test has been widely adopted in the photometric redshift literature \citep[e.g.,][]{qqUse1, qqUse2, qqUse3, qqUse4, qqUse5, qqUse6}. 

By definition, if the true redshift of a source lies within its $50\%$ CI, it will also lie within all CIs corresponding to larger percentages, as the $50\%$ CI will be a subset of these. It is therefore sufficient to measure only the threshold CI that just contains the true redshift. This will have one of the interval edges at the true redshift. This region can therefore be measured by drawing the horizontal line detailed above so that it intersects the posterior at the true redshift. The area $c$ corresponding to this interval is measured for each galaxy in the sample being tested. The cumulative distribution function (CDF) of these areas $\textrm{CDF}(c)$ can then be calculated. \cite{qqplot} note that for calibrated posteriors, the plot of this CDF against areas should be diagonal, i.e., $\textrm{CDF}(c) = c$. The deviation away from this line therefore measures how overconfident or underconfident the PDFs are.

\begin{figure}
	\includegraphics[width=\columnwidth]{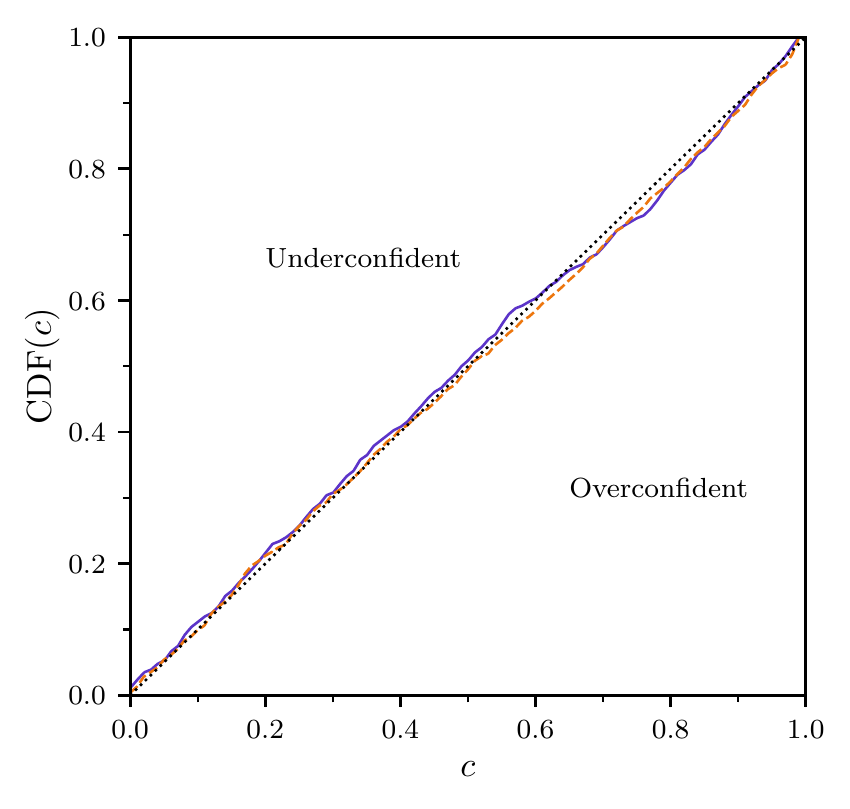}
	\caption{Plot showing the results of the posterior width test performed on posteriors obtained from our method on LSST-like simulated data. The solid purple line shows the results for the single-constituent posteriors, and the dashed orange line shows the results for the two-constituent posteriors. The black dotted line indicates the result where posteriors are calibrated, while lines that go above and below this indicate posteriors that are wider and narrower than calibrated posteriors respectively.}
	\label{fig:lsst-euclid-qq}
\end{figure}

A plot of this test for the LSST-like simulated data is shown in Fig.~\ref{fig:lsst-euclid-qq}. This figure shows that both the one- and two-constituent posteriors are approximately calibrated and their CIs can therefore be interpreted in a frequentist manner.

\begin{figure}
	\includegraphics[width=\columnwidth]{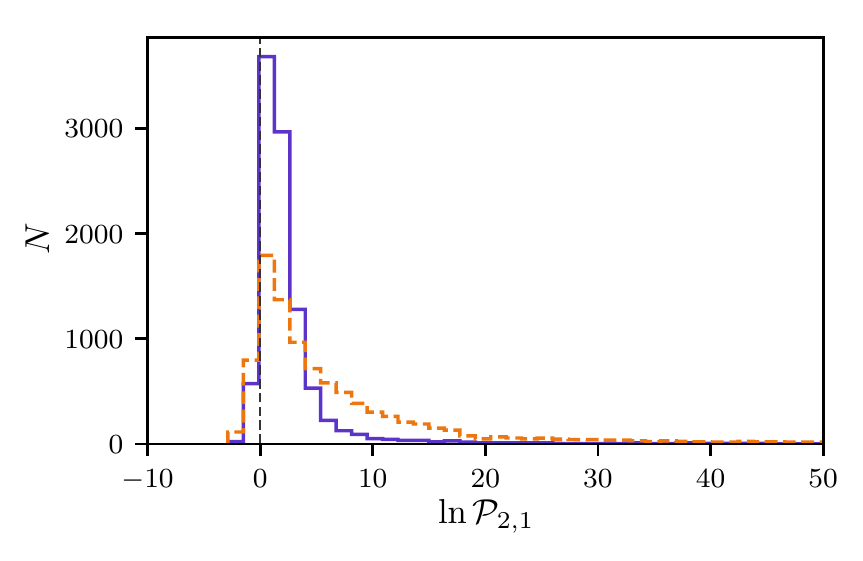}
	\caption{Histograms of the log of the relative probabilities for the blended and unblended models obtained using Bayesian model comparison on the simulated blended data. The solid purple histogram shows the result for the LSST-like survey, while the dashed orange histogram shows the result for the combined LSST-Euclid-like survey. The black dashed line indicates no preference for either the unblended or blended model. Larger values of $\mathcal{P}_{2,1}$ favour the blended model more.}
	\label{fig:lsst-bayes}
\end{figure}

Finally, Fig.~\ref{fig:lsst-bayes} shows the relative probability for the blended and unblended models $\mathcal{P}_{2, 1}$ calculated for the blended data of both simulated surveys. This quantity is calculated using the evidences derived in sections~\ref{sec:gmm-single-evd} and \ref{sec:gmm-blend-evd} using equation~\ref{eqn:model-select}. We assume a ratio of model priors of unity, i.e., we do not \textit{a priori} favour either the one- or two-constituent models. A blended source is then favoured when $\ln \mathcal{P}_{2, 1} > 1$. We find that the LSST-like survey identifies $92.4\%$ of blended sources, while the survey with additional infrared data identifies $89.3\%$.

\section{GAMA blended sources catalogue} \label{sec:gama-results}
In addition to the simulated observations presented in section~\ref{sec:sim-results}, we also test our method against real observations. To do this, we use data from the Galaxy And Mass Assembly (GAMA) survey~\citep{gamaData}, a spectroscopic survey of $>150 \, 000$ sources. Alongside this spectroscopy, these sources were also imaged in optical wavelengths by the Sloan Digital Sky Survey (SDSS)~\citep{sdssData} and in infrared wavelengths by the  VISTA Kilo-degree Infrared Galaxy (VIKING) Survey~\citep{vikingData}. \cite{gamaPhotometry} used this imaging data to create self-consistent, aperture-matched photometry in nine bands $u,g,r,i,z,Y,J,H,K$ for all sources within the GAMA survey. As a result, these sources have both high-quality photometry and accurate spectroscopic redshifts for training and testing our photometric redshift method. 

\cite{gamaBlends} used this data to spectroscopically identify blended sources in order to search for strong-lens candidates. The resulting GAMA blended sources catalogue contains blended photometry for $280$ sources, alongside the spectroscopic redshift of each constituent. We therefore use this catalogue to test the performance of our method on real observations of blended sources. To accompany this, we also randomly select two sets of $10000$ unblended sources for a training and test set.

\begin{figure}
	\includegraphics[width=\columnwidth]{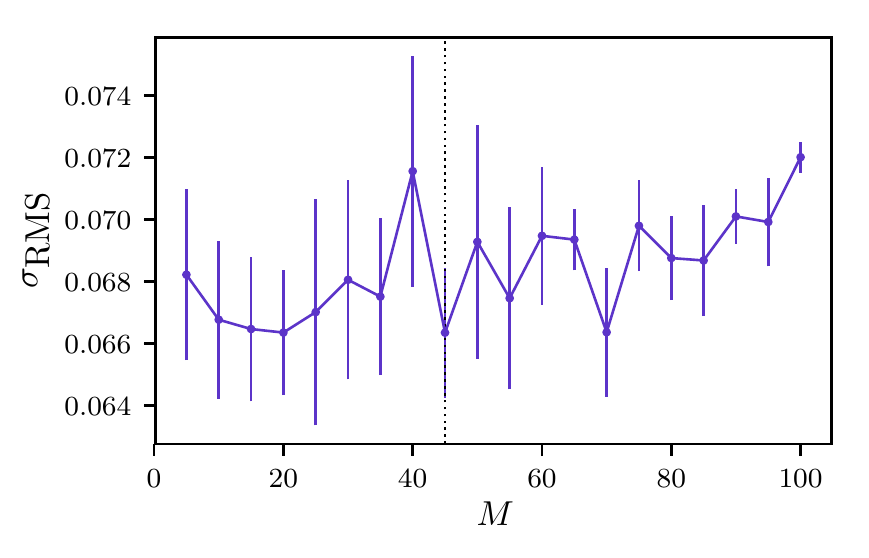}
	\caption{Results of the cross-validation for the GAMA blended sources catalogue data. The points show the RMS scatter averaged over the three folds, while the error bars show the error on the mean. We choose the number of components to be $N=45$, minimising the average RMS scatter as indicated by the dotted black line.}
	\label{fig:gama-crossVal}
\end{figure}

As for the simulated observations, we use 3-fold cross-validation to find the number of mixture components $N$ that minimises $\overline{\sigma_\textrm{RMS}}$ the RMS scatter averaged over all folds. The results of this are shown in Fig.~\ref{fig:gama-crossVal}. We find the minimum scatter when the number of mixture components is $N=45$, giving $\overline{\sigma_\textrm{RMS}}=0.066$. We therefore continue with a GMM of $45$ components fitted to the $10000$ unblended training sources.

\begin{figure}
	\includegraphics[width=\columnwidth]{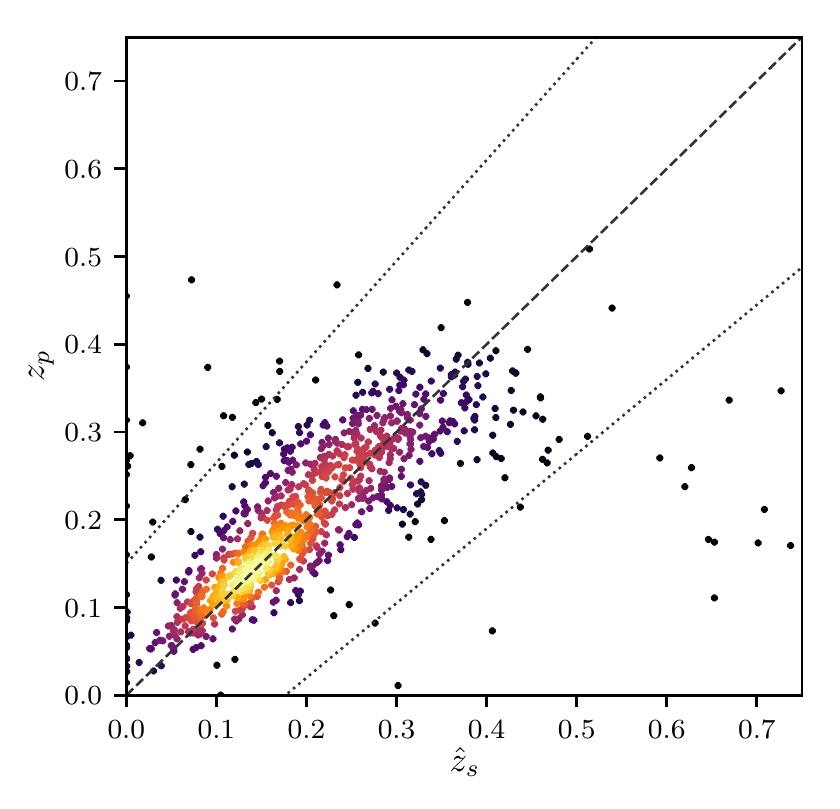}
	\caption{Plot showing the point-estimate results obtained from the GMM on the unblended GAMA data. The dashed line denotes $z_\textrm{p} = \hat z_\textrm{s}$, and the dotted lines indicate our outlier definition where $|z_\textrm{p} - \hat z_\textrm{s}| \geq 0.15 (1 + \hat z_\textrm{s})$. Points are coloured according to their density on the scatter plots to illustrate overplotting.}
	\label{fig:gama-single-scatter}
\end{figure}

We then compute point estimates of the single-constituent redshifts by averaging samples drawn from the posterior as before. A plot of this is shown in Fig.~\ref{fig:gama-single-scatter}. We find the RMS scatter to be $\sigma_{\textrm{RMS}}=0.067$, with $3.6\%$ of sources being outliers.

\begin{figure*}
	\includegraphics[width=\textwidth]{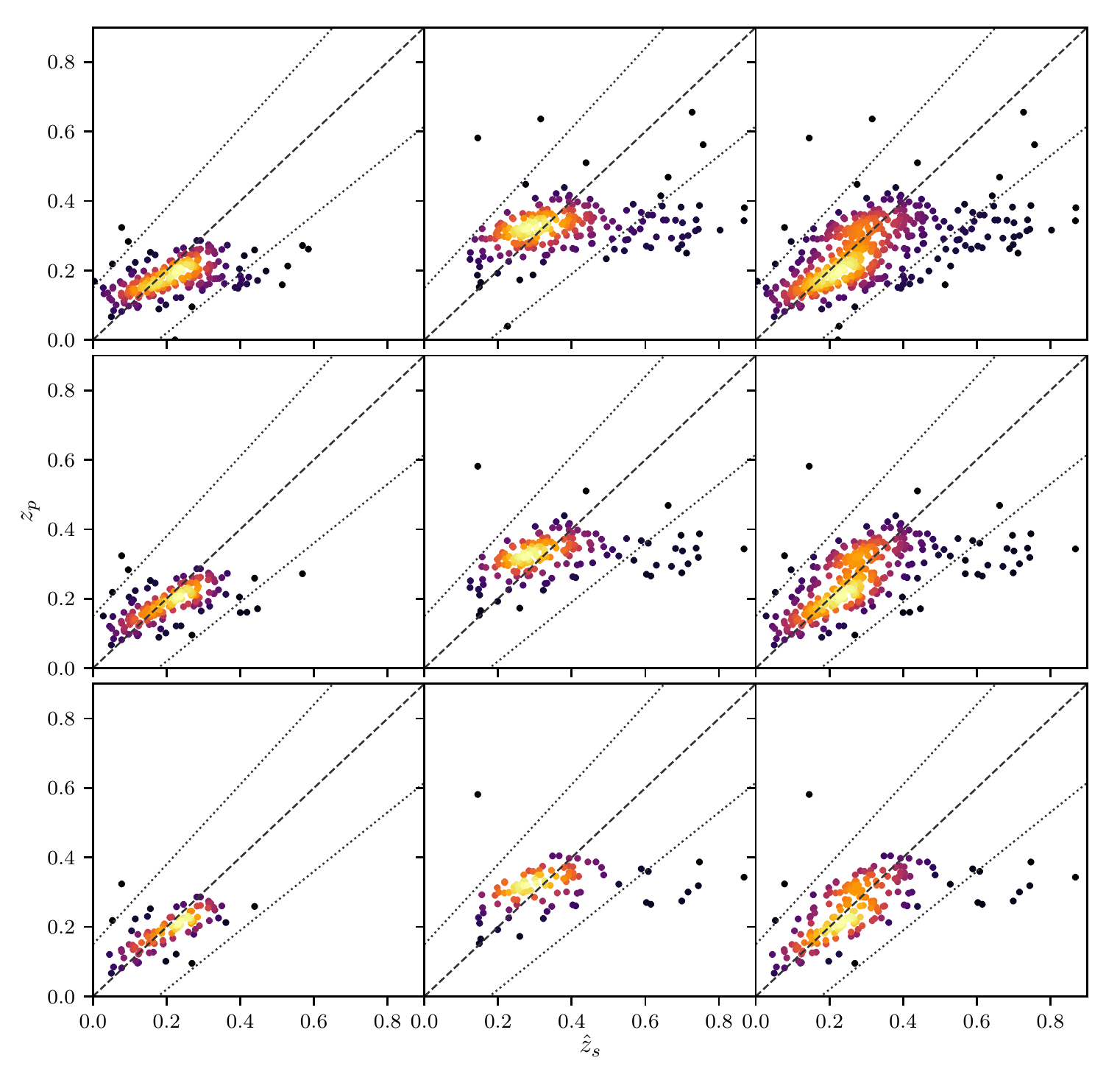}
	\caption{Plot showing the point-estimate results obtained from the GMM on the data from the GAMA blended sources catalogue, with various density ratio thresholds. The left column shows $z_{\textrm{p}, 1}$, the point estimate of the redshift for the lower-redshift constituent in each blended source. The centre column shows $z_{\textrm{p}, 2}$, corresponding to the higher-redshift constituent in each blended source. The right column combines both $z_{\textrm{p}, 1}$ and  $z_{\textrm{p}, 2}$. The top row shows the results for the full sample, while the centre and bottom rows have sources with expected density ratios less than $0.45$ and $0.8$ removed respectively. {where the expected density ratio is defined in equations~\ref{eqn:gmmz:expect-density-ratio} and~\ref{eqn:gmmz:thresh-density-ratio}. Imposing this density ratio threshold removes sources that are least well-represented in the training set, and so we would expect the results to improve as the threshold is increased. As indicated in the text, the summary statistics improve as expected by making these cuts. This can also be seen visually in this figure by comparing the lower two rows with the full sample in the top row.} The dashed lines denotes $z_\textrm{p} = \hat z_\textrm{s}$, and the dotted lines indicate our outlier definition where $|z_\textrm{p} - \hat z_\textrm{s}| \geq 0.15 (1 + \hat z_\textrm{s})$. Points are coloured according to their density on the scatter plots to illustrate overplotting.}
	\label{fig:gama-blend-scatter}
\end{figure*}

A scatter plot of the two-constituent point estimates is shown in Fig.~\ref{fig:gama-blend-scatter}. As in the simulated case, the blended results are noisier than the single-constituent case. We find the RMS scatter to be $\sigma_{\textrm{RMS}}=0.091$, and $10.8\%$ of sources to be outliers. 

\begin{figure*}
	\includegraphics[width=\textwidth]{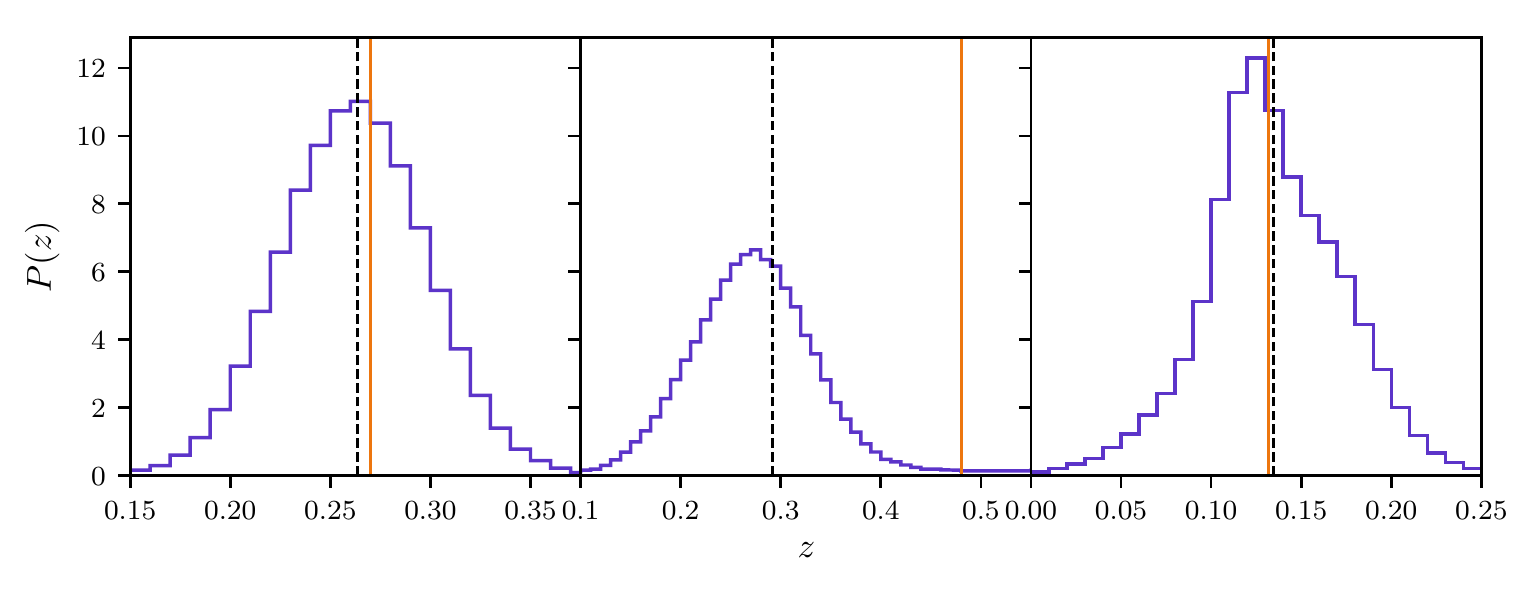}
	\caption{Plot showing three examples of single-constituent posteriors sampled using the GMM on the unblended GAMA data. The black dashed lines indicate the sample means we use to define the point estimates $z_\textrm{p}$. The true redshifts are indicated by the orange lines.}
	\label{fig:gama-single-post}
\end{figure*}

Examples of single-constituent posteriors are shown in Fig.~\ref{fig:gama-single-post}. Like the single-constituent posteriors conditioned on the simulated data, these distributions show a variety of shapes. However, the posteriors for the GAMA data are significantly less multimodal. This is likely because the GAMA sources are, on average, lower redshift than the simulated sources. The main cause of the bimodality in the simulated case is the colour-redshift degeneracy described in section~\ref{sec:gmm}, which low- and high-redshift sources to be confused. However, high redshifts are \textit{a priori} very unlikely here, as they do not appear in the training set. As a result, these higher redshift peaks are significantly disfavoured.

\begin{figure*}
	\includegraphics[width=0.32\textwidth]{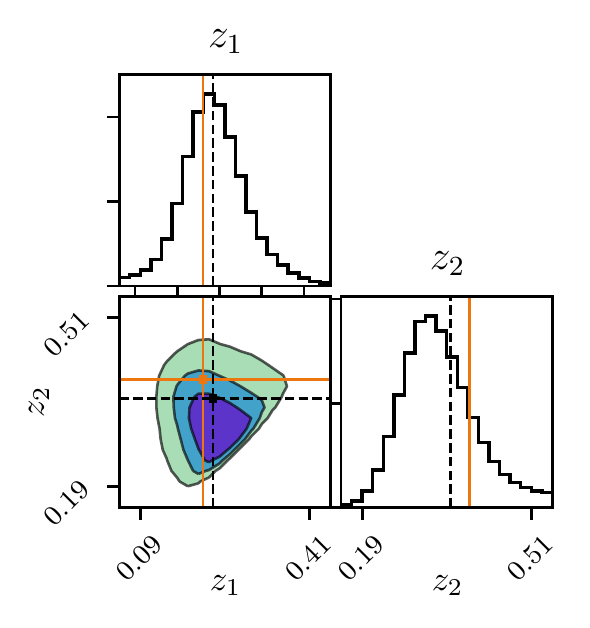}
	\includegraphics[width=0.32\textwidth]{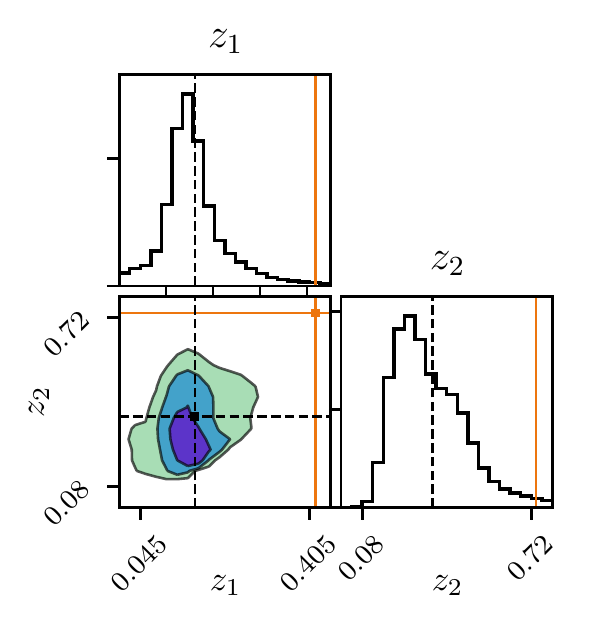}
	\includegraphics[width=0.32\textwidth]{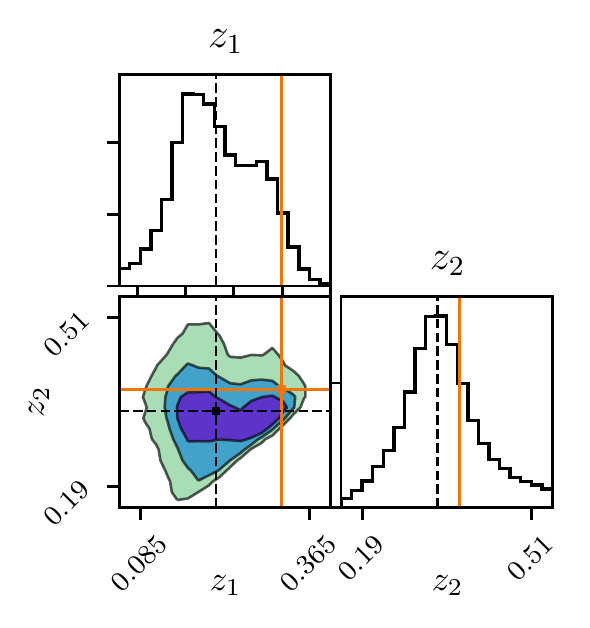}
	\caption{Plot showing three examples of two-constituent posteriors sampled using the GMM on data from the GAMA blended sources catalogue. The black dashed lines indicate the sample means we use to define the point estimates $z_\textrm{p}$. The true redshifts of each constituent are indicated by the orange lines.}
	\label{fig:gama-blend-post}
\end{figure*}

The same lack of multimodality is also exhibited in the blended posteriors conditioned on the GAMA data. Examples of these are shown in Fig.~\ref{fig:gama-blend-post}. These posteriors show a variety of non-Gaussian shapes as in the simulated case, with many of the marginal redshift distributions displaying long tails. The joint distribution in the left panel of Fig.~\ref{fig:gama-blend-post} also shows the hard cut resulting from the sorting condition $\pi(z_1, z_2)$, as the left panel of Fig.~\ref{fig:lsst-blend-post} does.

\begin{figure}
	\includegraphics[width=\columnwidth]{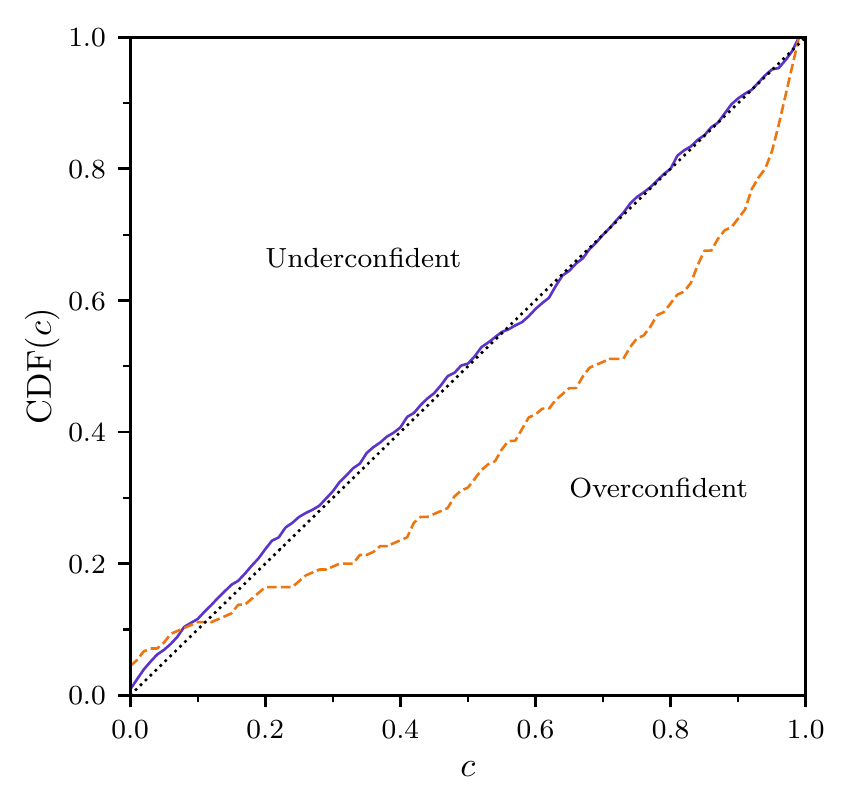}
	\caption{Plot showing the results of the posterior width test performed on posteriors obtained from our method on GAMA data. The solid purple line shows the results for the single-constituent posteriors, and the dashed orange line shows the results for the two-constituent posteriors. The black dotted line indicates the result where posteriors are calibrated, while lines that go above and below this indicate posteriors that are wider and narrower than calibrated posteriors respectively.}
	\label{fig:gama-qq}
\end{figure}

Fig.~\ref{fig:gama-qq} shows the plot testing the posterior widths for both the one- and two-constituent posteriors. As in the simulated case, the one-constituent posteriors are very close to being calibrated. However, the CDF for the two-constituent posteriors lies significantly below the diagonal, suggesting that the posteriors are overconfident, i.e., they are too narrow. As discussed above, while it is not guaranteed that Bayesian CIs provide frequentist coverage probabilities, this suggests that there are features on the flux-redshift relation of the blended constituents that are not captured by the model trained on the unblended training data. 

\begin{figure}
	\includegraphics[width=\columnwidth]{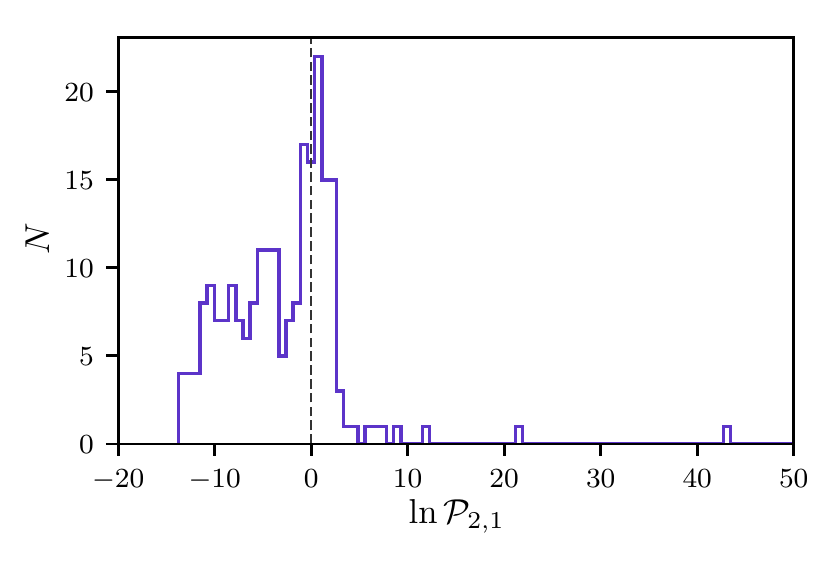}
	\caption{Histogram of the log of the relative probabilities for the blended and unblended models obtained using Bayesian model comparison on the blended GAMA data. The black dashed line indicates no preference for either the unblended or blended model. Larger values of $\mathcal{P}_{2,1}$ favour the blended model more.}
	\label{fig:gama-bayes}
\end{figure}

This interpretation is supported by Fig.~\ref{fig:gama-bayes} which shows the inferred relative probability of {sources from the blended sources catalogue} being blended and unblended $\mathcal{P}_{2, 1}$. Here, only $33.4\%$ of {blended }sources are {correctly} identified as blended by having $\mathcal{P}_{2, 1} >1$. While the redshifts are reasonably well-recovered, the Bayesian model selection will disfavour a more complicated model when the improvement in the fit is insufficient. As above, this suggests a difference between the blended and unblended constituents.

We can test for a difference between the blended and unblended constituents by incrementally removing sources where this difference is greatest and checking whether this leads to an improvement in the summary statistics. We therefore require a quantity to probe the representativeness of a given vector of fluxes. For this, we consider the density ratio
\begin{equation}
\mathcal{R}(\vct{F}) = \frac{P_\textrm{test}(\vct{F})}{P_\textrm{train}(\vct{F})} \,,
\end{equation}
where $P_\textrm{train}(\vct{F})$ is the density of fluxes is the training set, $P_\textrm{train}(\vct{F})$ is the density of fluxes is the test set and $\vct{F}$ is the flux vector at which both of these densities are evaluated. 

In order to estimate this ratio, we use the nearest-neighbour method of \cite{nnratio}. The method first considers the training set, and measures the hypervolume that contains the $n_\textrm{nei}$ nearest neighbours of a flux $\vct{F}$. The number of test-set samples $n_\textrm{test}(\vct{F})$ within that hypervolume centred on $\vct{F}$ is then counted. The estimate for the density ratio is then given as the ratio of these counts, i.e., 
\begin{equation}
\mathcal{R}(\vct{F}) \approx \frac{n_\textrm{nei}}{n_\textrm{test}(\vct{F})} \,.
\end{equation}

This nearest-neighbour method for estimating the density ratio was first presented in \cite{reweightMethod}, and was used to estimate the redshift distribution of a photometric galaxy sample by weighting spectroscopic galaxies. However, the accuracy of this method depends on $n_\textrm{nei}$, the number of neighbours considered. If $n_\textrm{nei}$ is too large, the density ratio is estimated over too large a volume, while an estimate where $n_\textrm{nei}$ is too small will be dominated by statistical errors. To this end, \cite{nnratio} present a model-selection method based on cross-validation to optimise $n_\textrm{nei}$.

As discussed throughout this paper, a complication of blended sources is that the flux of each constituent is not observed independently, only the blended combination. As a result, the destiny ratio must be evaluated using constituent fluxes sampled from the marginal posterior $P(\vct{F}_n | \hat{\vct{F}})$, where $\vct{F}_n$ is the flux of constituent $n$. As described in section~\ref{sec:gmm-blend-post}, this can be accomplished by sampling from the simplified posterior defined in equation~\ref{eqn:h-theta}, and rejecting samples that do not obey the boundary prior. The marginalisation over all redshifts and the flux of the other constituent can than be done by simply ignoring these elements of the sampled vectors. 

Given a set of $n_\textrm{F}$ flux samples $\{\vct{F}_n^i \pbar i = i \dots n_\textrm{F} \}$ from constituent $n$, we evaluate the density ratio $\mathcal{R}(\vct{F})$ for each sample and average the result to give the expectation value
\begin{equation}
\label{eqn:gmmz:expect-density-ratio}
\textrm{E} [\mathcal{R}(\vct{F})] 
\equiv \int \mathcal{R}(\vct{F}_n) P(\vct{F}_n | \hat{\vct{F}}) \drv \vct{F}_n
\approx
\frac{1}{n_\textrm{F}} \sum_{i} \mathcal{R}(\vct{F}_n^i) \,.
\end{equation}
This expectation value is the quantity we use to estimate the representativeness of blended constituents. This allows us to test for differences between the blended and unblended constituents. To do this, we keep sources in our sample only if the expectation of the density ratio for both of their constituents is over a threshold value $\mathcal{R}_\textrm{th}$, i.e., sources that obey
\begin{equation}
\label{eqn:gmmz:thresh-density-ratio}
\frac{\textrm{E} [\mathcal{R}(\vct{F}^i_n)]}{\textrm{max}(\textrm{E} [\mathcal{R}(\vct{F})])} \geq \mathcal{R}_\textrm{th}, \quad n \in \{1, 2\}
\,,
\end{equation}
where we have normalised the expectation values by $\textrm{max}(\textrm{E} [\mathcal{R}(\vct{F})])$, the maximum expectation value over both constituents of all sources.

\begin{figure}
	\includegraphics[width=\columnwidth]{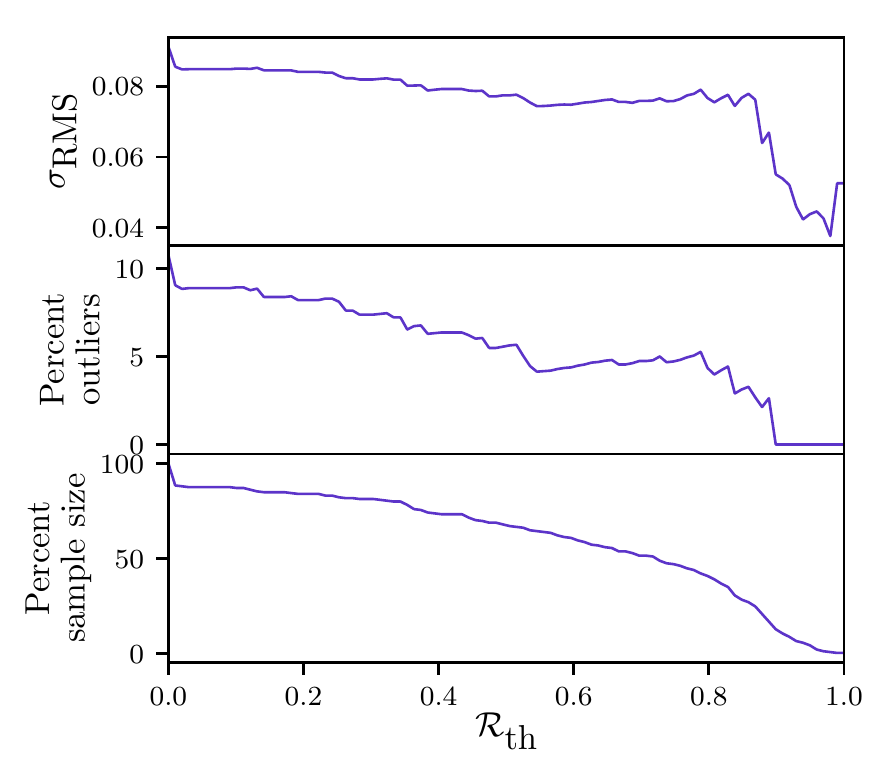}
	\caption{Plot showing the change in summary statistics for the GAMA blended sources as the density ratio threshold $\mathcal{R}_\textrm{th}$ is increased. The top panel shows the RMS scatter $\sigma_\textrm{RMS}$. The centre panel shows the percentage of sources that are outliers, defined as $|z_\textrm{p} - \hat z_\textrm{s}| \geq 0.15 (1 + \hat z_\textrm{s})$. The bottom panel shows the percentage of sources remaining from the original sample after the threshold has been applied.}
	\label{fig:gama-ratioCut_stats}
\end{figure}

Fig.~\ref{fig:gama-ratioCut_stats} shows the change in summary statistics as the threshold ratio is increased. As expected, the RMS scatter and number of outliers are both reduced as this ratio is increased, at the expense of more sources being removed from the sample. This effect can also be seen in the lower two rows of Fig.~\ref{fig:gama-blend-scatter}, where the effects of two different threshold values on the point estimates are compared with the unmodified results. When the threshold is set at $\mathcal{R}_\textrm{th} = 0.45$ as in the centre row, the RMS scatter has been reduced to $\sigma_\textrm{RMS}=0.078$, while the percentage of sources that are outliers has reduced to $5.97\%$. At this level, $70.7\%$ of sources remain in the sample. By increasing the threshold to $\mathcal{R}_\textrm{th} = 0.8$ as in the bottom row, the RMS scatter and percentage of outliers decrease to $\sigma_\textrm{RMS}=0.077$ and $4.34\%$ respectively. These are modest improvements over the less strict threshold, but come at the cost of leaving only $40.9\%$ of sources remaining in the sample.

These results demonstrate the importance of representative training sets. Differences between the training and test sets, often referred to as covariate shift, are a general problem for machine learning-based methods that obtain all of their information from the training set. A possible cause of differences here is that surveys select sources based on a magnitude cut, imparting selection effects on the sample. Since blended sources will be selected based on their total blended flux, blended constituents can be fainter than those that are unblended. The simulated sources presented in section~\ref{sec:sim-results} are selected in this way and so contain this effect. However, the intrinsic properties of galaxies vary with magnitude, meaning that the test set could contain faint constituents that have no corresponding examples in the test set. {Selection effects imparted by the selection criteria of sources in the blended sources catalogue, such as certain redshift differences being easier to select spectroscopically, are also not accounted for here.}

One solution to this problem is to improve the training set so that it is more representative. By including sources in the training set fainter than the magnitude limit of the test set, the model can learn the faint-end flux-redshift relation. The selection effects of blended sources could also be learned directly by training using a blended training set as described in section~\ref{sec:gmm-train}. However, as detailed above, assembling a representative blended training set in practice could be difficult. For the tests presented here, the GAMA blended sources catalogue contains far too few sources to be amenable to fitting in this way.

\section{Conclusions} \label{sec:conclusions}

Future galaxy surveys will observe to unprecedented depths in order to drive their increases in precision of cosmological constraints. However, these improvements to constraints on cosmological parameters will be accompanied by several new complications to the analysis. The increased number density of sources will increase both the number of sources that are blended and the total number of sources observed. 

This paper presents a photometric redshift method for blended sources based on Gaussian mixture models. Using these models, our method learns the flux-redshift distribution from a set of unblended training galaxies. This choice of model permits the derivation of posteriors that can be sampled efficiently, allowing the method to scale to large samples. By using Bayesian model selection techniques, this method can also infer the number of constituents within a blended sources efficiently.

{This work extends previous uses of GMMs in photometric redshift applications~\cite{xdQSOz} to the case of blended sources. It also extends the template-based} method to infer the redshifts of blended sources directly from their blended photometry first introduced in \cite{blendz}. The method described therein {relies on nested sampling for inference and so will not scale to the large sample sizes of future galaxy surveys such as LSST \citep{lsstSummary}.} The method presented in this paper is significantly faster, making it suitable for these upcoming surveys. {Many modern methods of photometric redshifts are machine learning-based, as training these methods on a representative training set can allow them to achieve very high accuracy and avoid the problems associated with small template sets. This paper extends the blended photometric redshift method of \cite{blendz} to this data-driven approach.}

The accuracy of all machine learning-based photometric redshift methods is dependent of the training set. Using training sets that are unrepresentative could result in redshift inferences that are biased and posterior distributions that are too narrow. In cases where unblended galaxies are not representative of individual components in a blended source, potentially as a result of selection effects, our method can generalise to learn the blended flux-redshift relation directly from blended training data. While this naturally accounts for differences between blended and unblended galaxies, it also increases the size of the required training set.

The method presented here represents a different approach to analysing blended sources than is currently used. Rather than separating blended observations into separate constituents, we infer the redshifts jointly for all constituents. As a result, our method naturally captures uncertainties and correlations which can be difficult to estimate for deblending-based analyses. This approach could be extended to other quantities of interest for cosmological analysis such as galaxy shapes by constructing forward models of source images. By doing this, correlations associated with blending can be propagated fully throughout the rest of the analysis, providing the best understanding of uncertainties on cosmological constraints.

\section*{Acknowledgements}
We thank Andrew Jaffe, Daniel Mortlock and Boris Leistedt for helpful discussions, {and the anonymous referee for many useful suggestions that have improved this paper.}
%
% Researchfish lists project reference as 1708326, googling that gives the following URL with a grant number on it
% https://gtr.ukri.org/projects?ref=studentship-1708326
DMJ acknowledges funding from STFC through training grant ST/N504336/1.
GAMA is a joint European-Australasian project based around a spectroscopic campaign using the Anglo-Australian Telescope. The GAMA input catalogue is based on data taken from the Sloan Digital Sky Survey and the UKIRT Infrared Deep Sky Survey. Complementary imaging of the GAMA regions is being obtained by a number of independent survey programmes including GALEX MIS, VST KiDS, VISTA VIKING, WISE, Herschel-ATLAS, GMRT and ASKAP providing UV to radio coverage. GAMA is funded by the STFC (UK), the ARC (Australia), the AAO, and the participating institutions. The GAMA website is http://www.gama-survey.org/.
Based on observations made with ESO Telescopes at the La Silla Paranal Observatory under programme ID 179.A-2004. 
%%%%%%%%%%%%%%%%%%%%%%%%%%%%%%%%%%%%%%%%%%%%%%%%%%



%\bibliographystyle{mnras}
%\bibliography{gmmzBib} % if your bibtex file is called example.bib

%\begin{thebibliography}{99}
%\end{thebibliography}

% Don't change these lines
\bsp	% typesetting comment
\label{lastpage}
\end{document}